\documentclass{natureprintstylev4}
\usepackage{astjnlabbrev-nature}
\usepackage{hyperref}
\usepackage[switch]{lineno}

\usepackage{amssymb}
\usepackage{amsmath}
\usepackage{mathtools}
\usepackage{gensymb}
\usepackage[svgnames]{xcolor}
\usepackage{enumitem}
\usepackage{textcomp}
\usepackage{subcaption}
\usepackage[numbers,sort&compress]{natbib}

\newcommand{\arcsec}{^{\prime\prime}}
\newcommand{\farcs}{.^{\prime\prime}}

\usepackage[normalem]{ulem}

\usepackage{epsfig}
\usepackage{graphicx}
\usepackage{longtable}
\usepackage{hyperref}
\hypersetup{
   colorlinks=true,
   linkcolor=blue,
   filecolor=green,
   citecolor=blue,
   urlcolor=blue
}

\usepackage[utf8]{inputenc}  
\usepackage[T1]{fontenc}     
\usepackage{textcomp}        

\DeclareUnicodeCharacter{223C}{$\sim$}   
\DeclareUnicodeCharacter{03BC}{$\mu$}    
\DeclareUnicodeCharacter{2272}{$\lesssim$} 
\DeclareUnicodeCharacter{2273}{$\gtrsim$} 
\DeclareUnicodeCharacter{03B1}{$\alpha$} 
\DeclareUnicodeCharacter{03B2}{$\beta$}  
\DeclareUnicodeCharacter{03BB}{$\lambda$}  
\DeclareUnicodeCharacter{2217}{$*$}  %

\newcommand{\hb}{\mbox{H$\beta$}}
\newcommand{\ha}{\mbox{H$\alpha$}}

\usepackage[dvipsnames]{xcolor}

\newcommand{\MgII}{Mg\,\textsc{ii}\,$\lambda\lambda2796,2803$}
\newcommand{\nad}{Na\,{\sc D}}
\newcommand{\OIII}{[O\,\textsc{iii}]}
\newcommand{\OIIIwl}{[O\,\textsc{iii}]$\,\lambda\lambda\,4960,5008$}
\newcommand{\OI}{O\,\textsc{i}\,$\lambda8446$}
\newcommand{\NII}{[N~\textsc{ii}]}
\newcommand{\HeIop}{He\,\textsc{i}\,$\lambda5877$}
\newcommand{\HeI}{He\,\textsc{i}\,$\lambda10830$}
\newcommand{\Pagamma}{Pa$\gamma$}

\newcommand{\FeII}{Fe\,\textsc{ii}}

\newcommand{\Lya}{Ly$\alpha$}

\newcommand{\Msunyr}{$\mathrm{M_\odot\,yr^{-1}}$}

\newcommand{\tablecomments}[1]{{\small #1}}

\usepackage[labelsep=endash]{caption}

\title{A quasar hatching from a buried red phase at $z=3.7$}

\author{
Zheng Ma$^{1,\dag}$,
Yongda~Zhu$^{1,\dag}$,
Zhiyuan Ji$^{1}$,
Eiichi Egami$^{1}$,
Marcia J.\ Rieke$^{1}$,
Xiaohui Fan$^{1}$,
Jianwei Lyu$^{1}$,
George H.\ Rieke$^{1}$,
Fengwu Sun$^{2}$,
Yang Sun$^{1}$,
George D.\ Becker$^{4}$,
Andrew J.\ Bunker$^{6}$,
Francesco D'Eugenio$^{7,8}$,
Xiangyu Jin$^{3}$,
Ignas Juodžbalis$^{7,8}$,
Weizhe Liu$^{1}$,
Roberto Maiolino$^{7,8,10}$,
Pierluigi Rinaldi$^{9}$,
Feige Wang$^{3}$,
Christopher N.\ A.\ Willmer$^{1}$,
Yunjing Wu$^{5}$,
Jinyi Yang$^{3}$,
Junyu Zhang$^{1}$,
Peixin Zhu$^{2}$
}

\begin{document}

\maketitle

\let\thefootnote\relax
\footnotetext{$^{\dag}$Emails: mazh@arizona.edu; yongdaz@arizona.edu}

\begin{affiliations}
\item Steward Observatory, University of Arizona, 933 North Cherry Avenue, Tucson, AZ 85721, USA
\item Center for Astrophysics $|$ Harvard \& Smithsonian, 60 Garden St., Cambridge, MA 02138, USA
\item Department of Astronomy, University of Michigan, 1085 S. University Ave., Ann Arbor, MI 48109, USA
\item Department of Physics \& Astronomy, University of California, Riverside, CA 92521, USA
\item Kavli Institute for the Physics and Mathematics of the Universe (WPI), The University of Tokyo Institutes for Advanced Study, The University of Tokyo, Kashiwa, Chiba 277-8583, Japan
\item Department of Physics, University of Oxford, Denys Wilkinson Building, Keble Road, Oxford OX1 3RH, UK
\item Kavli Institute for Cosmology, University of Cambridge, Madingley Road, Cambridge, CB3 0HA, UK 
\item Cavendish Laboratory, University of Cambridge, 19 JJ Thomson Avenue, Cambridge, CB3 0HE, UK 
\item Space Telescope Science Institute, 3700 San Martin Drive, Baltimore, Maryland 21218, USA 
\item Department of Physics and Astronomy, University College London, Gower Street, London WC1E 6BT, UK 
\end{affiliations}

\begin{abstract}
We present JADES-GS 209777, previously cataloged as CANDELS J033238.02$-$274626.2, hereafter “the Hatchling,” a red quasar at $z=3.711$. While the source has been reported in earlier deep-field catalogs, our multiwavelength analysis reveals a visible active nucleus still embedded in a dense gas- and dust-rich environment. Red quasar continua are often attributed to dust attenuation, including non-standard extinction curves, but the highly comprehensive multiwavelength data for this source provide direct constraints on the material being cleared. Using JWST/NIRSpec, NIRCam, MIRI, \textit{HST}, MUSE, \textit{Chandra}, ALMA, and VLA data, we detect broad emission lines and strong X-ray emission, showing that the active nucleus is at least partially exposed. We also detect H$\alpha$ and He\,{\sc i} absorption, indicating dense gas close to the nucleus. Kinematically disturbed O\,{\sc i}, Mg\,{\sc ii}, Na\,{\sc d}, and [O\,{\sc iii}] features, together with extended Ly$\alpha$ emission over $\gtrsim20~{\rm kpc}$, further show that multiphase gas is being accelerated from the nuclear region into the host-galaxy environment. The ALMA detection reveals strong dust emission, with the inferred infrared luminosity placing the system in the ULIRG regime. The continuum is red and sharply declining toward the rest-frame UV, resembling compact red AGNs, and may reflect extreme dust attenuation, gas reprocessing, possible BAL-like absorption, or a combination of these effects. Regardless of which mechanism dominates the continuum shape, the line diagnostics show that the visible nucleus remains partially obscured by nearby material. The Hatchling therefore represents a unique opportunity to explore a poorly known transition phase in which feedback is likely clearing an enshrouded quasar and allowing it to emerge toward a more unobscured active nucleus.
\end{abstract}



\bigskip

\section*{INTRODUCTION}

A central question in galaxy evolution is whether obscured and red quasars represent transitional stages through which rapidly growing black holes emerge as unobscured quasars. In merger-driven evolutionary scenarios, gas inflows trigger intense star formation and rapid black-hole growth in dust-enshrouded systems such as ultraluminous infrared galaxies (ULIRGs). Feedback from the accreting black hole may subsequently expel or redistribute the surrounding gas and dust, opening sightlines toward the nucleus and allowing the system to emerge as an optically visible quasar \citep{sanders_ultraluminous_1988, hopkins_unified_2006, hopkins_cosmological_2008}.

Here we report the Hatchling, a red quasar at $z=3.711$ in the JADES GOODS-South field. The source is covered by JWST/NIRSpec, JWST/NIRCam, JWST/MIRI, \textit{HST}, MUSE, \textit{Chandra}, ALMA, and VLA observations. This highly comprehensive dataset reveals broad permitted lines, an infrared excess and millimeter dust-continuum emission, absorption from dense and cool gas along nuclear sightlines, a broad blueshifted [O\,{\sc iii}] component, extended \Lya{} emission, and X-ray and radio counterparts. The Hatchling is therefore distinguished by the coexistence of a directly visible red Type-1 quasar with substantial dusty, gaseous, outflowing, and circumgalactic material. These properties are consistent with a quasar that is no longer fully buried but remains embedded in an incompletely cleared nuclear and galactic environment.

The significance of this source lies in the broader difficulty of identifying quasars in such an emerging phase. Red quasars have long been discussed as possible candidates: some may be otherwise normal Type-1 quasars viewed through dusty sightlines, whereas a subset has been proposed to represent quasars emerging from a dust-enshrouded phase \cite{glikman_first-2mass_2012, hoshi_evolutionary_2025}. This evolutionary interpretation is supported by observations associating red quasar colors with strong outflow signatures, including ionized and molecular winds \citep{glikman_luminous_2018, stacey_red_2022, vayner_powerful_2021}. 
However, red colors and broad permitted lines are not by themselves sufficient to identify an emergence phase. A more direct test requires simultaneous constraints on the visible nucleus, dust-reprocessed emission, dense gas along nuclear sightlines, multiphase outflows, and gas on larger scales. Such an observational test has been difficult because the required diagnostics are rarely available for the same source. It requires multiwavelength coverage of the rest-frame UV-optical continuum, dust emission, X-ray/radio activity, and host environment, together with moderate- to high-resolution spectroscopy that can separate broad lines, narrow lines, dense-gas absorption, and multiphase outflows. The Hatchling provides such a test: its broad multiwavelength coverage allows these diagnostics to be examined together in a single red Type-1 quasar during a phase when the nucleus is visible but its surrounding environment has not yet been fully cleared.

\section*{RESULTS}

\begin{figure*}[!ht]
    \centering
    \includegraphics[width=0.9\linewidth]{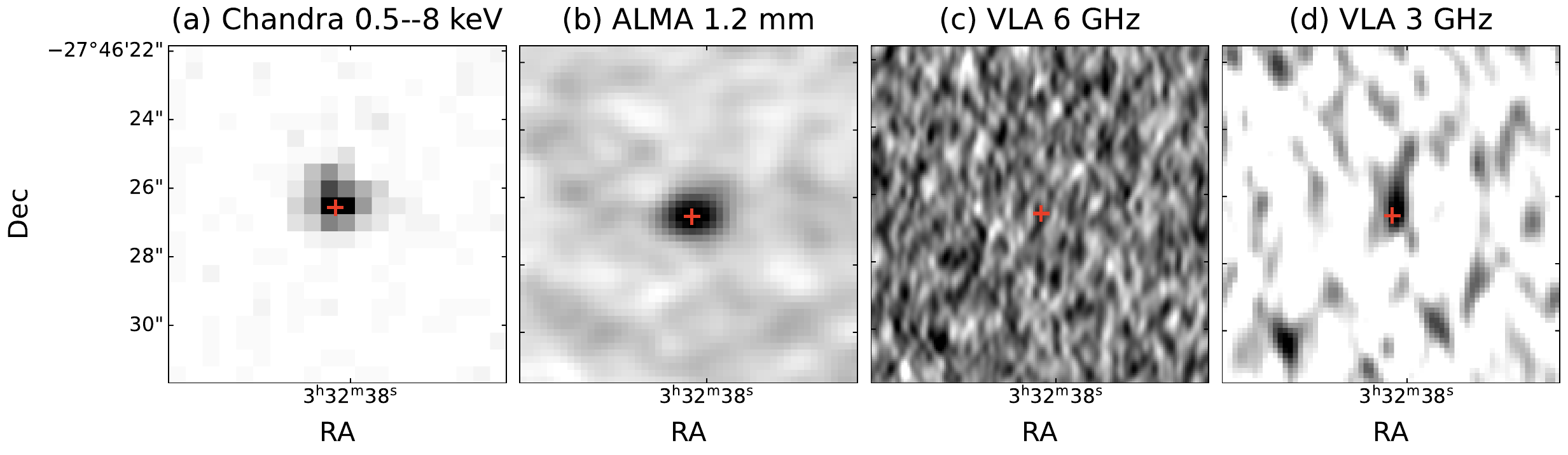}
    \includegraphics[width=0.9\linewidth]{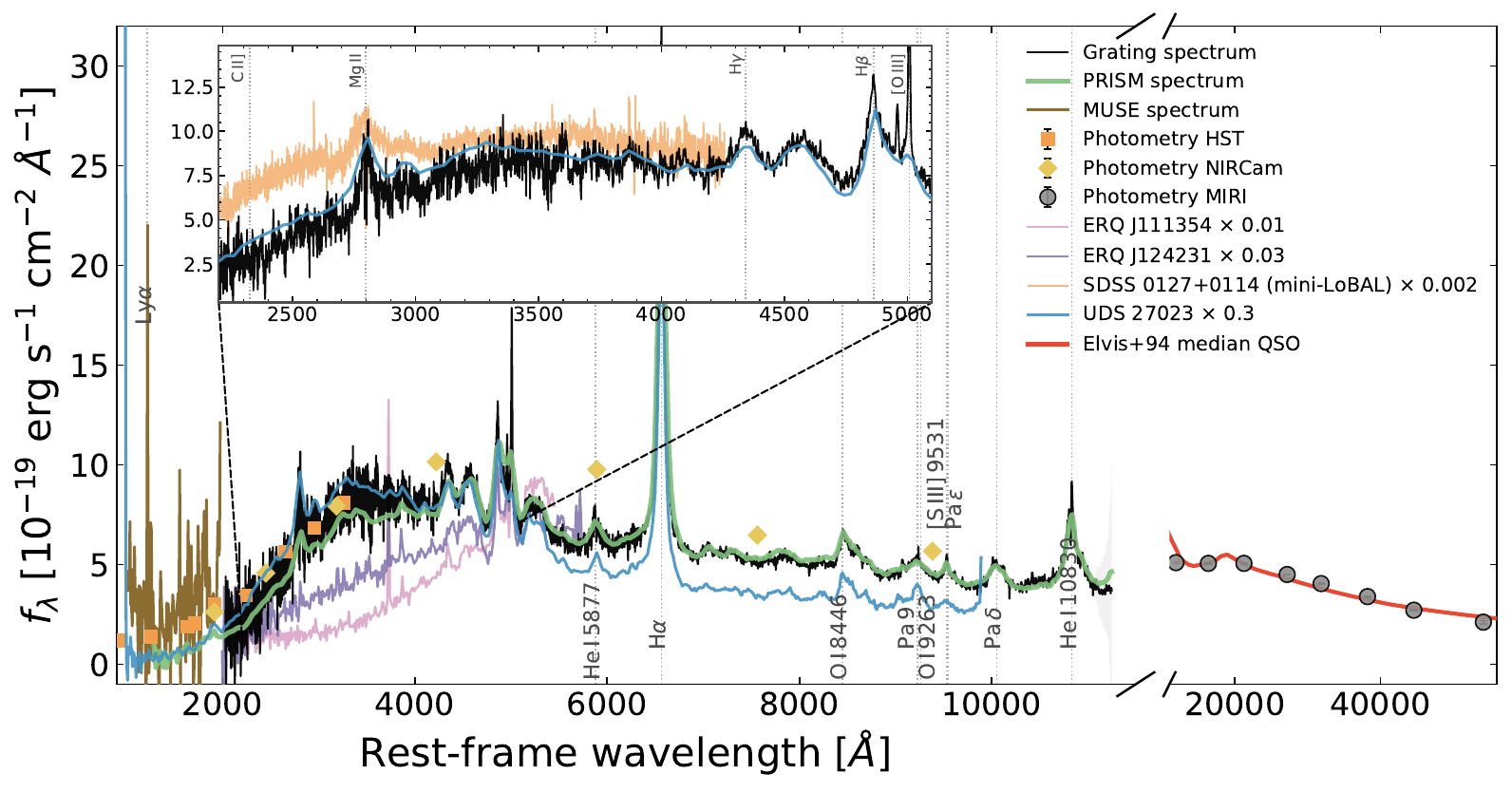}
    \caption{\textbf{Multiwavelength view and red rest-frame UV-optical continuum of the Hatchling.} Top panels show the \textit{Chandra} 0.5--8 keV, ALMA 1.2 mm, VLA 6 GHz, and VLA 3 GHz images, with red crosses marking the quasar position. The Hatchling is detected in X-rays, at ALMA 1.2 mm, and at VLA 3 GHz, but not significantly at VLA 6 GHz. The bottom panel shows the JWST/NIRSpec grating, JWST/NIRSpec PRISM, and MUSE spectra as the black, green, and brown solid curves, respectively. Photometry from \textit{HST}, JWST/NIRCam, and JWST/MIRI are shown as orange squares, yellow diamonds, and gray circles, respectively. The inset presents an enlarged view of the NIRSpec grating spectrum over rest-frame 2200--5100~\AA{}, highlighting the UV emission and absorption features and the \FeII{} pseudo-continuum. The Hatchling shows a broad rest-frame UV-to-optical bump peaked near 4000~\AA{}, similar to the red continua of enshrouded AGN, which may reflect dust attenuation, thermal-like reprocessing, BAL-like absorption, or a combination of these effects, while also displaying direct AGN signatures across X-ray, radio, and broad-line emission. For comparison, the pink and purple curves show two extremely red quasars selected from SDSS \cite{ross_extremely_2015}, the orange curve shows the reddened, strong \FeII{}-emitting mini-LoBAL quasar SDSS J0127+0114 \cite{hall_unusual_2002}, while the blue curve shows UDS 27023, a red quasar whose reddening has been attributed to small dust grains \cite{li_steep-extinction_2026}. All comparison spectra have been scaled for clarity. The red solid curve shows the near infrared SED of the majority of mature AGNs \cite{elvis_atlas_1994, lyu_dust-deficient_2017}, which provides a good match to the infrared SED of the Hatchling.}
    \label{fig:spectrum}
\end{figure*}

\begin{figure*}[!ht]
    \centering
    \includegraphics[width=0.9\linewidth]{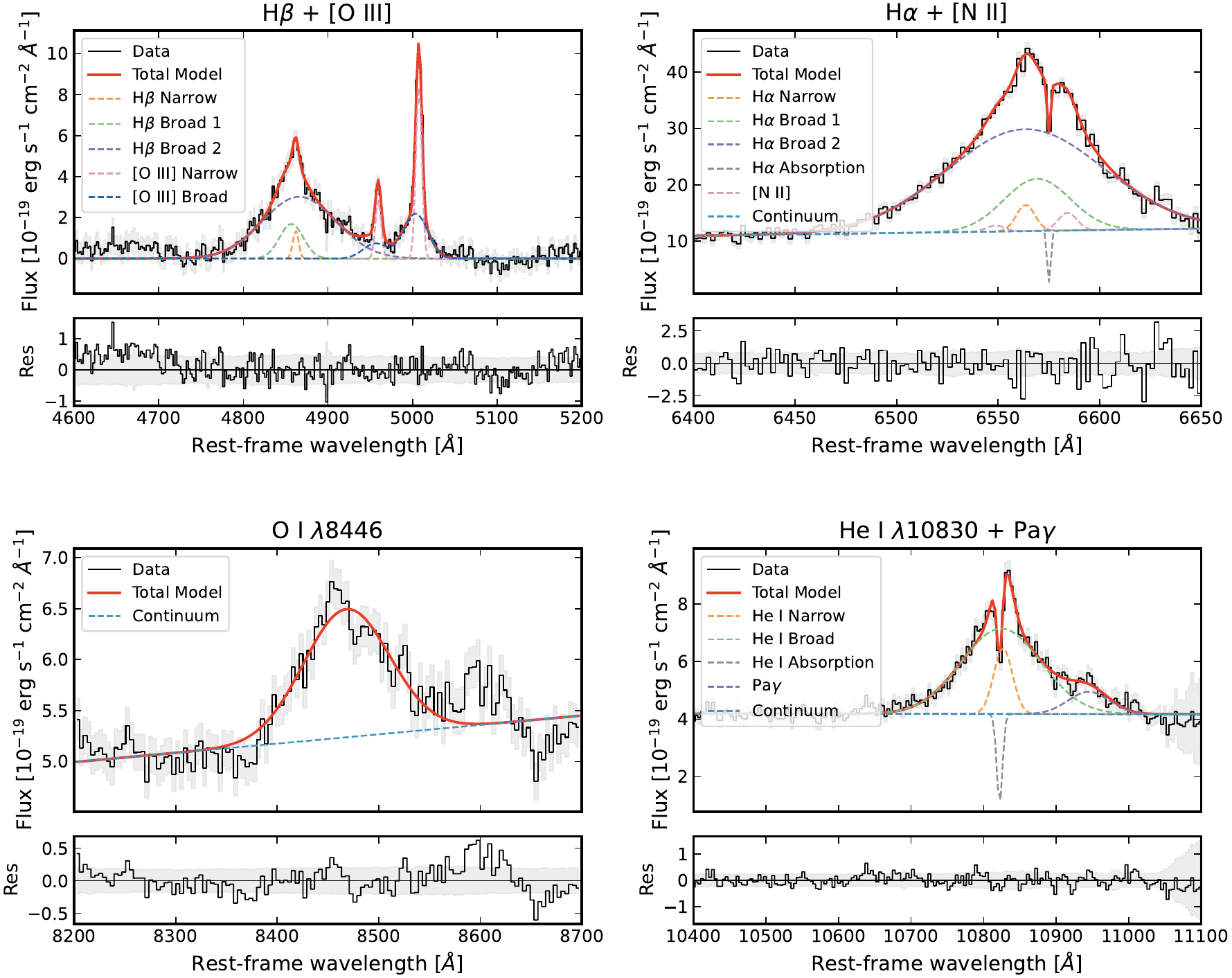}
    \caption{
    Dense gas and ionized outflows in the emerging nucleus. Line profiles of \hb, \OIII, \OI, and \HeI\ are measured from the NIRSpec grating spectra, while \ha\ is measured from the NIRCam grism spectrum. Broad components are required to reproduce the profiles of \hb, \ha, \OI, \HeI, and \Pagamma, while \OIII\ shows a broad blueshifted component tracing ionized outflowing gas. Absorption features are detected on top of the \ha\ and \HeI\ emission profiles, indicating dense gas along the line of sight. These features suggest that the nucleus is already visible, but remains embedded in dense, kinematically disturbed gas.
    }
    \label{fig:line_decomp}
\end{figure*}

\begin{figure*}[!ht]
    \centering
    \includegraphics[width=0.8\linewidth]{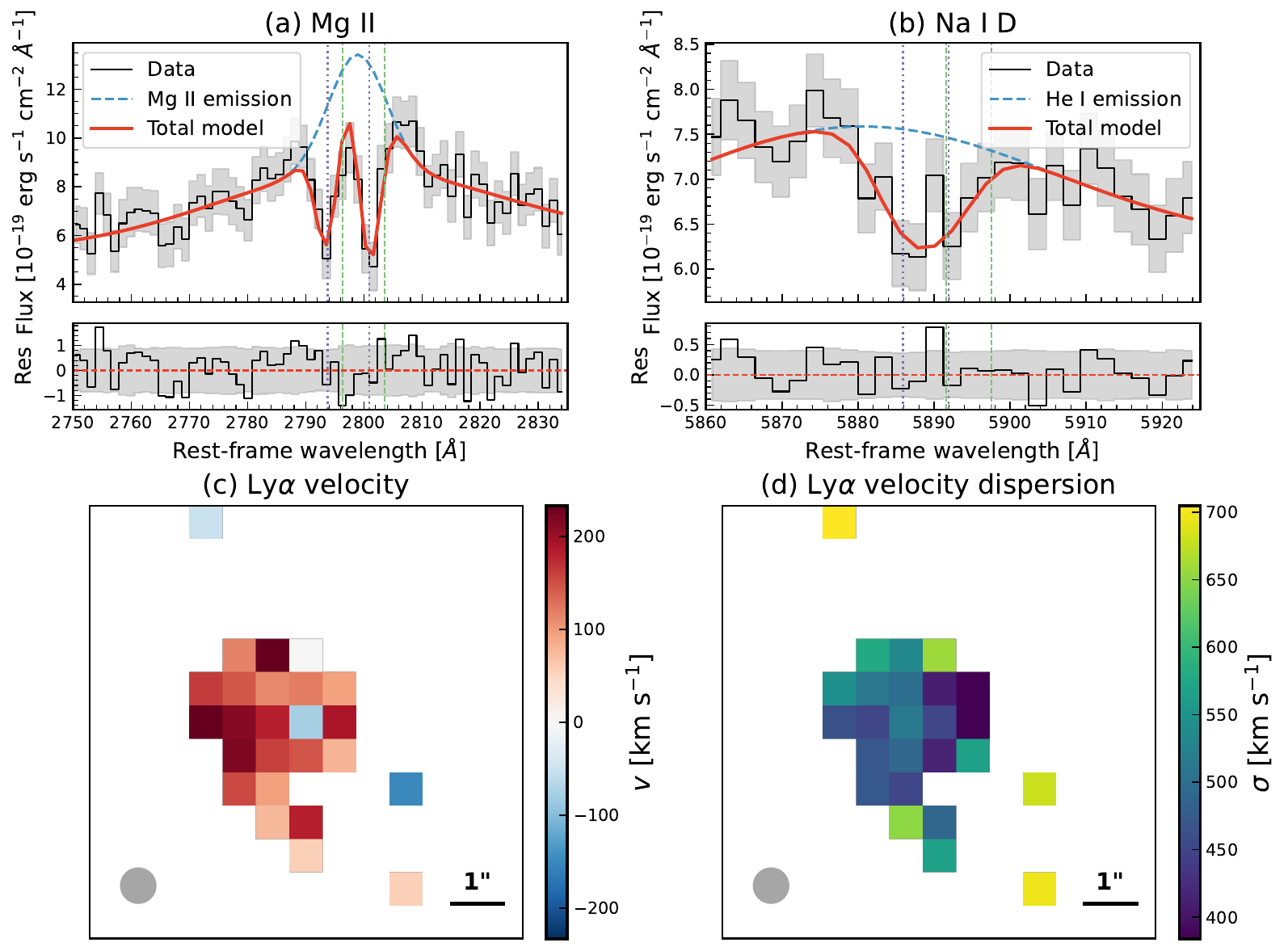}
    \caption{
    Multiphase clearing of the buried nucleus. The top panels show the Mg\,{\sc ii} (left) and Na\,{\sc i}\,{\sc D} (right) absorption profiles. The lower panels show the \Lya\ velocity map (left) and velocity-dispersion map (right). Because the \Lya\ detection has a low signal-to-noise ratio (S/N), we use maps rebinned by $3\times3$ spatial pixels and show only pixels with ${\rm S/N}>2$. The gray circle shows the MUSE PSF with FWHM $=$ 0.66$\arcsec$.
    These absorption features trace cool outflowing gas, while the extended \Lya\ halo reveals kinematically complex gas on $\sim20$ kpc scales. }
    \label{fig:cool_abs}
\end{figure*}

\begin{figure*}[!ht]
    \centering
    \includegraphics[width=0.8\linewidth]{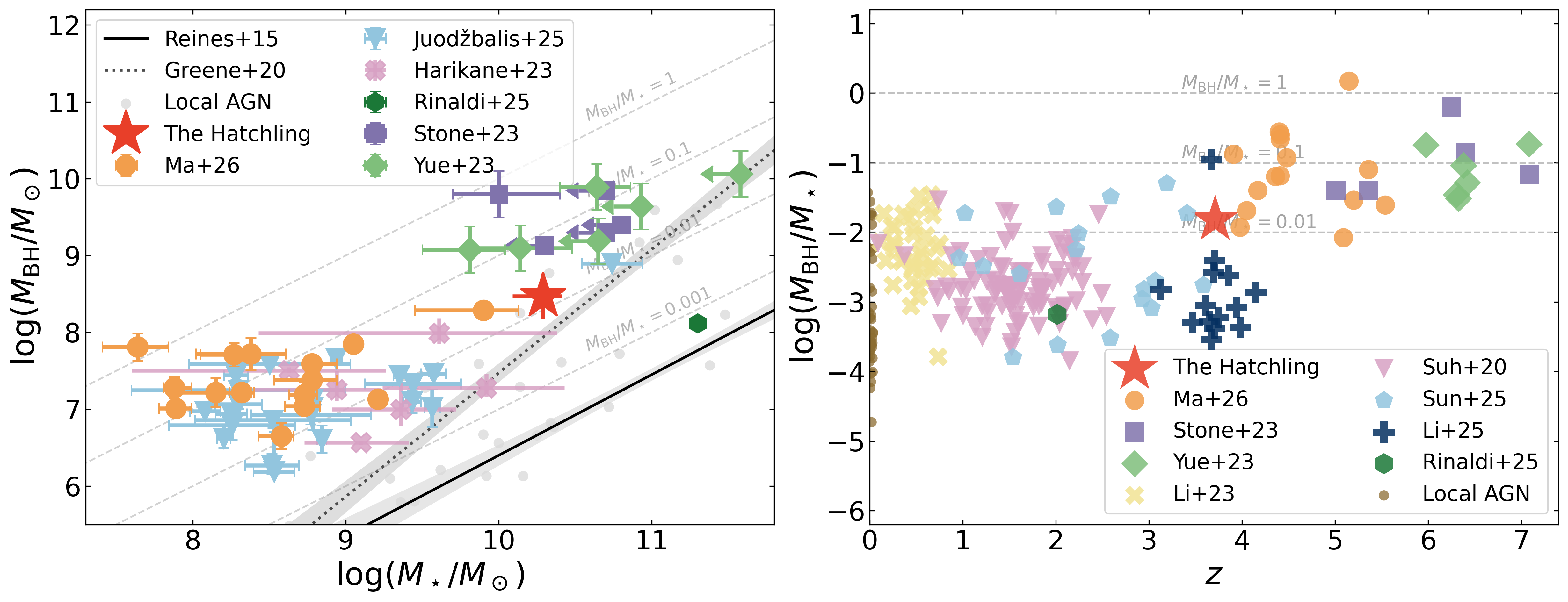}
    \caption{
    An emerging quasar in an intermediate black-hole growth regime.
    \textbf{Left:} Black-hole mass versus host-galaxy stellar mass. The Hatchling is marked by the red star. We compare it with faint AGNs from \cite{ma_undermassive_2026} (orange circles), \cite{harikane_jwstnirspec_2023} (pink crosses), and \cite{juodzbalis_jades_2025} (blue triangles), as well as high-redshift ($z\sim6$) quasars from \cite{stone_undermassive_2024} (purple squares) and \cite{yue_eiger_2024} (green diamonds). The local relations from \cite{reines_relations_2015} and \cite{greene_intermediate-mass_2020} (the relation for "all, limits") are shown as the black solid and dotted lines, respectively. The Hatchling lies above the local $M_{\rm BH}$--$M_\star$ relation, but below the most extreme overmassive quasars at high redshift. We also show the Saguaro as a dark-green hexagon \cite{rinaldi_beyond_2025}, which may represent a later evolutionary stage than the Hatchling.
    \textbf{Right:} Redshift evolution of $M_{\rm BH}/M_\star$. The Hatchling is again marked by the red star. We compare it with local AGNs from \cite{reines_relations_2015} (brown circles), low-redshift quasars from \cite{li_sloan_2023} (yellow crosses), broad-line AGNs from \cite{suh_no_2020} (pink downward triangles) and \cite{sun_no_2025} (sky blue pentagons), the Saguaro from \cite{rinaldi_beyond_2025} (dark-green hexagon), and high-redshift ($z>3$) AGNs from \cite{li_prevalent_2025} (navy blue plus signs), \cite{ma_undermassive_2026} (orange circles), \cite{stone_undermassive_2024} (purple squares), and \cite{yue_eiger_2024} (green diamonds). The Hatchling occupies an intermediate regime: its black hole lies above the local $M_{\rm BH}$--$M_\star$ relation, but it is not as extreme as the most overmassive high-redshift quasars or LRDs.}
    \label{fig:mbh_mstar}
\end{figure*}

\section*{Multiwavelength view of a red quasar at \mbox{\boldmath $z$\,$=$\,3.711}}

Figure~\ref{fig:spectrum} shows the multiwavelength view of the Hatchling. The NIRSpec grating and NIRCam grism spectra exhibit broad permitted emission lines and narrower forbidden lines, confirming the presence of an active nucleus. The source is matched to the 7 Ms CDF-S catalog source ID 715, with a catalog observed-frame full-band flux of $f_{\rm 0.5-7\,keV}=1.82\times10^{-15}$~erg~cm$^{-2}$~s$^{-1}$ (${\rm S/N}=29$) \cite{luo_chandra_2017}. At our systemic redshift of $z=3.711$, this corresponds to an observed-band luminosity of $L_{\rm 0.5 - 7\,keV,obs}\approx2.3\times10^{44}$~erg~s$^{-1}$. Assuming a power-law spectrum with $\Gamma=1.8$, the apparent rest-frame 0.5 -- 7 keV luminosity is $L_{\rm 0.5 - 7\,keV}^{\rm app}\approx1.7\times10^{44}$~erg~s$^{-1}$, well above the $L_{\rm X,int}\geq3\times10^{42}$~erg~s$^{-1}$ AGN-classification criterion used in the 7 Ms CDF-S catalog. The source is detected in the ALMA 1.2 mm image with $S_{1.2\,\rm mm}=163.2\pm10.1~\mu$Jy (${\rm S/N}=16.2$). The total infrared luminosity inferred by SED fitting gives $\log_{10}(L_{\rm IR}/L_\odot)=12.53\pm0.16$ \cite{boogaard_alma_2020, gonzalez-lopez_alma_2020, aravena_alma_2020}, above the conventional ULIRG threshold of $L_{\rm IR}>10^{12}\,L_\odot$ \citep{sanders_ultraluminous_1988}. Assuming the infrared luminosity is powered by star formation, it corresponds to an SFR of about 300 M$_\odot$~yr$^{-1}$, for a Kroupa IMF \citep{kennicutt_star_2012}: see {\it Methods}. At 3 GHz, the source is detected with a peak flux density of $S_{\rm 3GHz,peak}=1.98\pm0.10~\mu{\rm Jy~beam^{-1}}$ (${\rm S/N}=19.5$). The radio-to-optical ratio is $R_{4400} \sim 1$, assuming $S_\nu \propto \nu^{-0.7}$, which is consistent with a radio-quiet AGN classification ($R_{4400} < 10$) \cite{kellermann_vla_1989}. It is not significantly detected at 6 GHz; forced photometry at the JWST position gives $S_{\rm 6GHz,peak}=0.41\pm0.31~\mu{\rm Jy~beam^{-1}}$ (${\rm S/N}=1.3$) \cite{lyu_agn_2022,alberts_completing_2020}. The X-ray detection and visible broad permitted lines show that the active nucleus is at least partially exposed, while the ALMA detection indicates that dust remains present in the system, with an inferred dust mass of $M_{\rm d}\approx(2.1\,$--$\,9.9)\times10^7\,M_\odot$ for $T_{\rm d}=25\,$--$\,55\,{\rm K}$, or $M_{\rm d}\approx4.6\times10^7\,M_\odot$ for $T_{\rm d}=35\,{\rm K}$ (see {\it Methods}).


We observe a red continuum extending from the rest-frame UV to the optical ($\sim$\,2000--6000\,\AA), with a turnover at $\sim$\,4000\,\AA.  We will examine its physical origin in {\em Discussion} and {\em Methods}. We also show the near infrared (rest-frame) measurements and their fit with a ``normal'' near infrared SED, implying that the circumnuclear torus of heated dust is in place in the Hatchling.

\section*{Dense gas remains along the line of sight}

The NIRSpec spectrum reveals typical AGN signatures, including broad permitted emission lines. Figure~\ref{fig:line_decomp} presents the line-fitting results for \hb{}, \OIIIwl{}, \ha{}, \OI{} and \HeI{}. The fitting procedure and line measurements are described in \textit{Methods}, with the corresponding measurements listed in Table~\ref{tab:line_measurements}. The \ha{} profile is decomposed into one narrow component and two broad components, with broad-line full widths at half maximum (FWHM) of $\simeq2600$ and $\simeq5900~{\rm km~s^{-1}}$. Similarly, the \hb{} profile is decomposed into one narrow component and two broad components, with broad-line FWHM values of $\simeq2000$ and $\simeq6300~{\rm km~s^{-1}}$. These broad Balmer components demonstrate that at least part of the broad-line region (BLR) is directly visible \citep[e.g.,][]{yanagisawa_scaling_2026}.

However, absorption features superposed on the \ha{} and \HeI{} emission indicate that dense, kinematically complex gas remains along the line of sight \cite{hutchings_balmer_2002, zhang_reddening_2017}. The \ha{} profile from the NIRCam grism spectrum requires a redshifted absorption component with a velocity shift of $v\simeq+475~{\rm km~s^{-1}}$ and ${\rm FWHM}\simeq93~{\rm km~s^{-1}}$. Producing Balmer absorption requires a substantial population of hydrogen in the $n = 2$ level, generally implying relatively dense gas and/or a sufficiently large column density with efficient \Lya{} trapping. The redshifted \ha{} absorption is therefore consistent with dense gas moving inward along the line of sight. The \HeI{} profile also requires an absorption component superposed on the emission, with a blueshift of $\simeq -310~{\rm km~s^{-1}}$ and ${\rm FWHM}\simeq259~{\rm km~s^{-1}}$ (Table~\ref{tab:line_measurements}). Because the metastable \HeI{} transition can produce detectable absorption over a broader range of gas densities and column densities than Balmer absorption, the blueshifted component may instead trace a lower-density or lower-column outflow. The opposite velocity shifts thus indicate that the two absorption features arise from kinematically distinct gas components, plausibly dense inflowing gas in \ha{} and outflowing gas in \HeI{}.

The \OI{} line provides an additional tracer of dense, partially neutral gas near the nucleus. The line is fitted with a broad component whose centroid is redshifted relative to the systemic wavelength by $\simeq+700~{\rm km~s^{-1}}$ and whose width is ${\rm FWHM}\simeq3535~{\rm km~s^{-1}}$, indicating kinematically disturbed low-ionization gas (Table~\ref{tab:line_measurements}). \OI{} can be enhanced by Ly$\beta$ fluorescence in dense, high-column, partially neutral gas, and is commonly associated with the low-ionization BLR or dense circumnuclear material \cite{juodzbalis_jades_2025, geris_jades_2026}. Together with the \ha{} and \HeI{} absorption, which reveal dense gas along the line of sight to the broad-line nucleus, the broad O\,{\sc i} emission shows that the Hatchling contains a visible broad-line nucleus that remains at least partially covered by dense gas.

\section*{Multiphase outflows from ionized to cool gas}

The spectrum reveals outflowing gas in multiple phases. The [O\,{\sc iii}] $\lambda\lambda4960,5008$ complex shows a broad blueshifted component with ${\rm FWHM}\simeq2039~{\rm km~s^{-1}}$ and a velocity shift of $\simeq-286~{\rm km~s^{-1}}$ (Fig.~\ref{fig:line_decomp}; Table~\ref{tab:line_measurements}). This component traces fast ionized gas likely driven by the active galactic nucleus (AGN) or nuclear starburst. Its large width and blueshift indicate that ionized material is being expelled from the central region.

Cooler gas also participates in the outflow. Figure~\ref{fig:cool_abs} shows Mg\,{\sc ii} and Na\,{\sc d} absorption, both blueshifted by $\simeq -280$--$-290~{\rm km~s^{-1}}$ and with ${\rm FWHM}\simeq500~{\rm km~s^{-1}}$ (Table~\ref{tab:line_measurements}). These absorption features trace lower-ionization material commonly associated with cool outflows in galaxies and AGNs. Their presence in the Hatchling shows that the feedback process is multiphase, involving fast ionized gas, dense partially ionized gas, and cooler absorbing material.

We estimate mass outflow rates for the main observed phases (see \textit{Methods}). The broad \OIII{} component gives an ionized outflow rate of $\dot{M}_{\rm out}=6.4^{+0.8}_{-0.7}$~\Msunyr{}. The \MgII{} and \nad{} absorption lines give cooler-gas outflow rates of $\dot{M}_{\rm out}=15.3^{+4.8}_{-4.3}$~\Msunyr{} and $\dot{M}_{\rm out}=15.1^{+1.0}_{-0.8}$~\Msunyr{}, respectively. Summing over these tracers gives a representative multiphase outflow rate of $\dot{M}_{\rm out}\sim37$~\Msunyr{}.  These rates lie within the range of Na D-traced neutral outflows measured at 0.6~$<$~$z$~$<$~4, but are modest compared with the median rate of $\sim\,80\,$\Msunyr{} found for star-forming galaxies \cite{sun_census_2026}. They correspond to a mass-loading factor of $\eta\,\sim\,0.05 - 0.3$, indicating that the observed cool neutral phase is being expelled more slowly than the host is forming stars. This value is an order-of-magnitude estimate, since the different tracers may sample overlapping or geometrically distinct gas components and depend on assumptions about density, geometry, and covering factor, as discussed in \textit{Methods}. 



\section*{A disturbed Ly$\alpha$ halo on circumgalactic scales}

Disturbed gas is also detected outside the unresolved nucleus. MUSE observations reveal extended Ly$\alpha$ emission on a projected scale of $\sim20$ kpc (Fig.~\ref{fig:cool_abs}). The integrated Ly$\alpha$ profile has a non-parametric FWHM of $\sim$\,670 km s$^{-1}$, indicating kinematically broad circumgalactic emission (see \textit{Methods}).

We present the Ly$\alpha$ velocity and dispersion maps in Figure~\ref{fig:cool_abs}. Because Ly$\alpha$ is resonantly scattered, its spatial distribution and apparent velocity structure are shaped by gas kinematics as well as radiative-transfer effects, and therefore cannot be interpreted as a simple map of the underlying gas motion. Nevertheless, the Ly$\alpha$ emission is spatially extended, indicating the presence of gas on galactic scales. Together with the \OIII{}, \HeI{}, O\,{\sc i}, \MgII{}, and \nad{} diagnostics, which are similar to metal absorption tracers of enriched gas at high redshift \citep[e.g.,][]{zhu_early_2026,nakane_jwst_2026,keerthi_vasan_g_jwst_2026}, this suggests that complex gas kinematics are present across a wide range of spatial scales, from the central AGN environment to the extended halo.

\section*{An intermediate black-hole growth regime}
We estimate the host-galaxy stellar mass by fitting the image-decomposed host photometry, following the procedure of \cite{ma_undermassive_2026}, and derive a single-epoch black-hole mass estimate from the broad \ha{} emission line. The host-galaxy context of the Hatchling is shown in Figure~\ref{fig:mbh_mstar}. In the $M_{\rm BH}$--$M_\star$ plane, the Hatchling has $M_{\rm BH}/M_\star\sim0.01$ (see \textit{Methods}), placing it above the local relation \citep[e.g.,][]{reines_relations_2015, greene_intermediate-mass_2020} but below the most overmassive high-redshift AGNs and little red dot (LRD) systems at $z\sim5$. This offset indicates that the black hole is massive relative to the observed stellar component of the host, consistent with a scenario in which black hole growth precedes the full assembly of the host galaxy.

The redshift comparison in Fig.~\ref{fig:mbh_mstar} provides additional context. Recent work finds little or no strong evolution in typical $M_{\rm BH}/M_\star$ ratios over $1<z<4$ \citep{sun_no_2025}, whereas higher-redshift JWST-selected low-mass AGNs and LRDs include systems with substantially elevated $M_{\rm BH}/M_\star$ ratios \citep[e.g.,][]{ma_undermassive_2026, juodzbalis_jades_2025, harikane_jwstnirspec_2023}. Luminous high-redshift quasars can also show apparent positive offsets from the local relation, although these offsets are strongly affected by luminosity selection, host-mass measurement uncertainties, and intrinsic scatter \cite{stone_undermassive_2024, yue_eiger_2024, sun_m_-m_rm_2025}. The Hatchling may therefore represent an intermediate growth stage: the black hole has already reached a relatively high mass, while continued stellar-mass assembly in the host could move the system toward the lower $M_{\rm BH}/M_\star$ ratios observed in more typical lower-redshift AGNs. This interpretation remains suggestive, however, because the comparison is affected by selection effects, uncertainties in host-galaxy decomposition, and systematic uncertainties in single-epoch black-hole masses.

At the same time, SED fitting based on image-decomposed photometry gives a recent star formation rate, averaged over the last 30 Myr, of ${\rm SFR}_{30}=54^{+45}_{-33}$~\Msunyr{} (see \textit{Methods}). This is significantly less than the infrared-based estimate of ${\rm SFR}\approx300^{+136}_{-93}~M_\odot~{\rm yr^{-1}}$, a common situation among ULIRGs, which probably arises because the dust can compete successfully for ionizing photons in their extreme environments \citep{alberts_calibrating_2026}. These values  indicate that the host is still rapidly building its stellar mass. The black hole is already massive and driving feedback, while continued stellar-mass growth could move the system closer to the scaling relations observed at lower redshift. The Saguaro at $z \sim 2$ may represent a later stage of this track, because while its embedded nucleus shows a reddened spectrum, the host galaxy is already developed \cite{rinaldi_beyond_2025}.

The offset from the local $M_{\rm BH}-M_{\star}$ scaling relation, together with the active star formation, supports a picture in which the Hatchling is observed after substantial black-hole growth but while its host galaxy is still building up toward the lower-redshift scaling relations.

\section*{DISCUSSION}

\begin{figure*}
    \centering
    \includegraphics[width=0.8\linewidth]{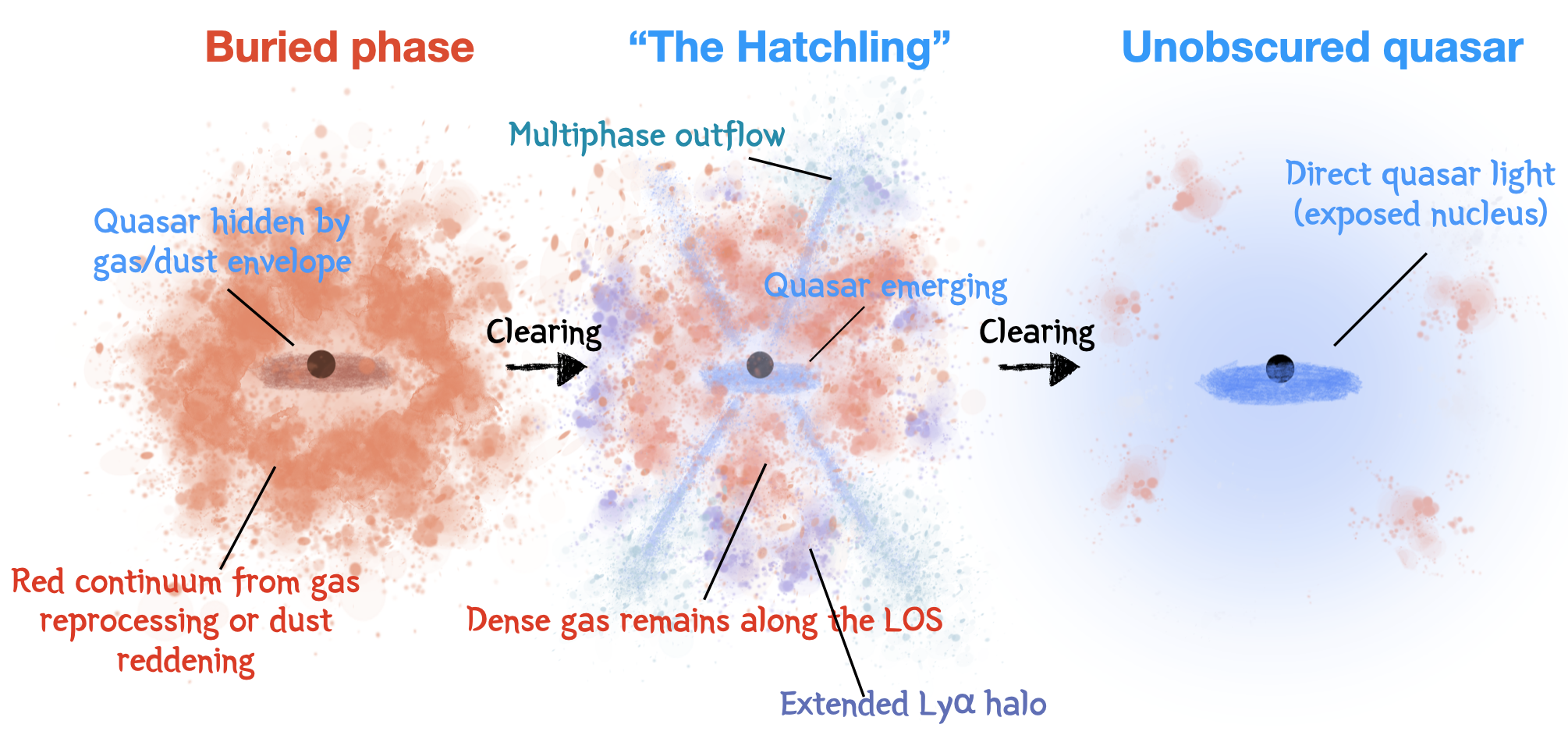}
    \caption{Schematic illustration of the proposed evolutionary interpretation for the Hatchling. In the buried phase, the quasar is largely hidden by dense gas and/or dust, producing a red continuum through dust attenuation, thermal-like reprocessing, BAL-like absorption, or a combination of these effects. In the Hatchling, the nucleus appears to be emerging as AGN-driven outflows begin to disrupt the surrounding gas and dust, plausibly creating low-opacity channels through which direct quasar light can escape. Dense gas nevertheless remains along some nuclear sightlines, producing absorption signatures. Continued feedback may clear the surrounding gas and drive the system toward an unobscured quasar phase, where direct quasar light dominates and only residual gas remains.}
    \label{fig:cartoon}
\end{figure*}

\section*{A partially exposed red quasar undergoing multiphase clearing}

The Hatchling is very red in the rest-frame 2000 - 3000 \AA ~range (see Figure~\ref{fig:spectrum}), where direct AGN emission emerges through a dense circumnuclear medium. Broad Balmer emission, X-ray emission, and radio emission show that the central engine is already visible. H$\alpha$ and He\,{\sc i} absorption, strong O\,{\sc i} $\lambda8446$ emission, Mg\,{\sc ii} and Na\,{\sc d} absorption, broad [O\,{\sc iii}], and the disturbed Ly$\alpha$ halo show that dense, multiphase gas remains from the nuclear region to circumgalactic scales. This combination identifies the Hatchling as a partially exposed red quasar.

The rest-frame UV-optical continuum shape of the Hatchling allows multiple explanations that are degenerate and difficult to distinguish observationally: standard dust laws do not recover a normal unobscured quasar spectrum, while small-grain dust attenuation, thermal-like reprocessing, absorption resembling that seen in broad absorption line (BAL) quasars \cite{hall_unusual_2002}, or some combination of these effects can provide phenomenological descriptions (see \textit{Methods}). We therefore do not use the continuum shape alone to define the physical state of the source. Instead, the key evidence for an enshrouded emerging nucleus comes from the coexistence of a visible broad-line AGN with dense-gas absorption and multiphase gas kinematics.

The line diagnostics constrain the clearing process across multiple gas phases. Broad [O\,{\sc iii}] traces fast ionized gas accelerated from the central region. Mg\,{\sc ii} and Na\,{\sc d} absorptions trace cooler outflowing material \citep[e.g.,][]{sun_extreme_2025}. H$\alpha$ and He\,{\sc i} absorptions mark dense gas along some lines of sight to the broad-line region. O\,{\sc i} $\lambda8446$ traces dense gas in the low-ionization broad-line region or in a dense reprocessing medium. The Ly$\alpha$ halo shows that the disturbed gas extends to projected scales of $\sim20$ kpc. These features point to a dense multiphase structure being accelerated and redistributed by AGN feedback.

Figure~\ref{fig:cartoon} illustrates this physical picture. Direct AGN radiation has begun to escape through lower-opacity channels, while dense gas remains along other directions. Continued feedback may reduce the covering fraction of this material and move the system toward a more unobscured quasar state. In more evolved ultraluminous quasars, multiphase outflows are observed after the central engine has largely cleared its immediate surroundings \citep[e.g.,][]{zhu_potential_2026}. The Hatchling appears to capture an earlier stage of this clearing process.

\section*{Is the Hatchling a bridge between red AGNs and mature quasars?}
The Hatchling may provide a bridge between compact red AGNs and more exposed quasars. It combines characteristics commonly associated with both compact red AGNs and more exposed quasars. Its strongly reddened continuum and dense circumnuclear gas resemble those of embedded red AGNs, whereas its broad permitted lines and X-ray emission demonstrate that the central engine is already directly visible across multiple wavelengths. At the same time, the presence of multiphase outflows suggests that the surrounding gas is being actively dispersed. This coexistence of residual obscuration, an exposed broad-line nucleus, and multiphase outflows makes the Hatchling a plausible transitional object between compact red AGNs and mature, unobscured quasars.

The host-galaxy context places this transition in a broader growth sequence. The Hatchling has $M_{\rm BH}/M_\star\sim0.01$, lies above the local $M_{\rm BH}$--$M_\star$ relation, and has a recent star-formation rate from total infrared luminosity of ${\rm SFR} \sim  300 $~\Msunyr{}, larger than the multiphase outflow rate of $\dot{M}_{\rm out}\sim37$~\Msunyr{}. At $z = 3.71$ and $\rm{log}$$(M_\star/M_\odot) = 10.29$, commonly used star-forming main-sequence relations predict $\rm{SFR}_{MS} = 74$~\Msunyr{} \cite{speagle_highly_2014}. This would therefore place the Hatchling approximately 0.6 dex, or a factor of four, above the main sequence. The host is therefore still rapidly building its stellar mass while feedback is already removing gas from the nuclear region. Nearby companions at similar redshift, discussed in 
\textit{Methods}, may trace an overdense environment or filamentary structure, providing a potential baryon reservoir for continued host-galaxy growth.

The Hatchling therefore captures a poorly explored but critical stage in quasar emergence: the accreting black hole is already visible, feedback is clearing the surrounding gas and dust, and the host galaxy is still assembling around an overmassive nucleus. Current samples of red quasars and compact red AGNs are still limited by selection, and follow-up spectroscopy \citep[e.g.,][]{juodzbalis_jades_2026} may still miss similar transition-phase objects, particularly if the transition occurs quickly. Systematic searches for such sources, followed by high-resolution rest-UV to optical spectroscopy, will be needed to determine how common this phase is among the red AGN population, and to help constrain its duration.

\begingroup
\setlength{\tabcolsep}{5pt}
\begin{table*}
\centering
\caption{Emission- and absorption-line measurements}
\label{tab:line_measurements}
\begin{tabular}{llcccc}
\hline
Line & Component & Flux & $v_{\rm shift}$ & FWHM & EW$_{\rm rest}$ \\
 &  & ($10^{-19}$ erg s$^{-1}$ cm$^{-2}$) & (km s$^{-1}$) & (km s$^{-1}$) & (\AA) \\
\hline
H$\beta$ & narrow & $11.5_{-4.3}^{+4.2}$ & $-51_{-11}^{+11}$ & $489_{-14}^{+8}$ & $1.9_{-0.7}^{+0.7}$ \\
H$\beta$ & broad 1 & $58.6_{-18.4}^{+31.9}$ & $-352_{-167}^{+129}$ & $2084_{-458}^{+695}$ & $9.8_{-3.1}^{+5.3}$ \\
H$\beta$ & broad 2 & $326.8_{-30.9}^{+20.8}$ & $191_{-128}^{+164}$ & $6274_{-363}^{+501}$ & $54.3_{-5.2}^{+3.5}$ \\
{[O III] $\lambda4960$} & narrow & $24.3_{-1.6}^{+1.4}$ & $-51_{-11}^{+11}$ & $489_{-14}^{+8}$ & $3.8_{-0.2}^{+0.2}$ \\
{[O III] $\lambda5008$} & narrow & $73.1_{-4.7}^{+4.2}$ & $-51_{-11}^{+11}$ & $489_{-14}^{+8}$ & $11.5_{-0.7}^{+0.7}$ \\
{[O III] $\lambda4960$} & broad & $26.3_{-2.1}^{+2.2}$ & $-290_{-85}^{+88}$ & $2033_{-182}^{+220}$ & $4.1_{-0.3}^{+0.3}$ \\
{[O III] $\lambda5008$} & broad & $79.1_{-6.4}^{+6.7}$ & $-290_{-85}^{+88}$ & $2033_{-182}^{+220}$ & $12.5_{-1.0}^{+1.0}$ \\
H$\alpha$ & narrow & $64.5_{-6.5}^{+6.8}$ & $-52_{-0}^{+0}$ & $489_{-0}^{+0}$ & $10.0_{-1.0}^{+1.1}$ \\
H$\alpha$ & broad 1 & $703.0_{-66.5}^{+66.4}$ & $-51_{-23}^{+23}$ & $2598_{-119}^{+119}$ & $109.1_{-10.7}^{+11.3}$ \\
H$\alpha$ & broad 2 & $1190.2_{-60.5}^{+57.7}$ & $88_{-40}^{+40}$ & $5866_{-210}^{+240}$ & $184.9_{-9.5}^{+8.9}$ \\
{[N II] $\lambda6549$} & narrow & $11.6_{-1.8}^{+1.8}$ & $-52_{-0}^{+0}$ & $489_{-0}^{+0}$ & $1.8_{-0.3}^{+0.3}$ \\
{[N II] $\lambda6585$} & narrow & $34.4_{-5.3}^{+5.3}$ & $-52_{-0}^{+0}$ & $489_{-0}^{+0}$ & $5.4_{-0.8}^{+0.8}$ \\
H$\alpha$ & absorption & $-18.8_{-7.9}^{+8.5}$ & $475_{-12}^{+12}$ & $92_{-51}^{+31}$ & $0.5_{-0.1}^{+0.2}$ \\
O I $\lambda8446$ & emission & $133.8_{-8.1}^{+8.8}$ & $694_{-87}^{+93}$ & $3534_{-217}^{+242}$ & $25.5_{-1.6}^{+1.8}$ \\
He I $\lambda10830$ & narrow & $70.5_{-25.7}^{+57.7}$ & $-170_{-70}^{+82}$ & $814_{-238}^{+556}$ & $16.8_{-6.2}^{+14.5}$ \\
He I $\lambda10830$ & broad & $411.7_{-26.7}^{+22.6}$ & $-262_{-107}^{+114}$ & $3528_{-323}^{+953}$ & $98.4_{-6.8}^{+6.3}$ \\
Pa$\gamma$ & emission & $55.8_{-24.5}^{+30.3}$ & $110_{-222}^{+197}$ & $1976_{-779}^{+876}$ & $13.4_{-5.9}^{+7.6}$ \\
He I $\lambda10830$ & absorption & $-31.1_{-7.9}^{+6.3}$ & $-308_{-17}^{+18}$ & $260_{-35}^{+26}$ & $3.5_{-0.6}^{+0.7}$ \\
Mg II & narrow & $65.9_{-44.7}^{+66.7}$ & $98_{-215}^{+472}$ & $1234_{-591}^{+495}$ & $11.6_{-7.9}^{+11.8}$ \\
Mg II & broad & $180.0_{-18.6}^{+17.6}$ & $105_{-218}^{+220}$ & $6128_{-771}^{+887}$ & $31.7_{-3.3}^{+3.3}$ \\
Mg II $\lambda2796$ & absorption & $-28.5_{-3.3}^{+2.9}$ & $-277_{-21}^{+23}$ & $501_{-39}^{+46}$ & $2.5_{-0.3}^{+0.3}$ \\
Mg II $\lambda2803$ & absorption & $-41.7_{-3.6}^{+3.8}$ & $-277_{-21}^{+23}$ & $501_{-39}^{+46}$ & $3.2_{-0.3}^{+0.3}$ \\
Mg II doublet & absorption total & $-70.4_{-4.7}^{+4.6}$ & $-277_{-21}^{+23}$ & $501_{-39}^{+46}$ & $5.7_{-0.4}^{+0.4}$ \\
He I 5877 & emission & $89.2_{-13.8}^{+16.9}$ & $145_{-169}^{+150}$ & $3058_{-480}^{+612}$ & $14.4_{-2.3}^{+2.9}$ \\
Na I D2 $\lambda5891$ & absorption & $-8.3_{-4.4}^{+4.0}$ & $-289_{-126}^{+68}$ & $485_{-193}^{+194}$ & $1.1_{-0.5}^{+0.6}$ \\
Na I D1 $\lambda5897$ & absorption & $-6.4_{-5.4}^{+3.3}$ & $-289_{-126}^{+68}$ & $485_{-193}^{+194}$ & $0.9_{-0.4}^{+0.7}$ \\
Na I D doublet & absorption total & $-15.5_{-3.3}^{+3.0}$ & $-289_{-126}^{+68}$ & $485_{-193}^{+194}$ & $2.1_{-0.4}^{+0.4}$ \\
\hline
\end{tabular}
\vspace{0.5em}
\begin{minipage}{0.98\textwidth}
\footnotesize
\textit{Notes.} Columns:
(1) Spectral line.
(2) Kinematic component used in the spectral decomposition.
(3) Integrated line flux, in units of $10^{-19}$ erg s$^{-1}$ cm$^{-2}$; absorption fluxes are reported as negative flux deficits.
(4) Velocity shift relative to the adopted systemic redshift.
(5) Full width at half maximum (FWHM).
(6) Rest-frame equivalent width.

For H$\beta$ and [O\,{\sc iii}], equivalent widths are measured relative to the fitted continuum model from \textsc{PyQSOFit}. For the remaining emission lines, equivalent widths are measured relative to the local power-law continuum adopted in each line fit. For absorption lines, equivalent widths are measured relative to the unabsorbed local model. Reported uncertainties correspond to the 16th--84th percentile ranges of the posterior samples. Total equivalent widths are computed from the posterior distribution of the summed absorption components and therefore are not necessarily equal to the sum of the separately reported posterior medians. For the narrow H$\alpha$ and [N {\sc ii}] components, the velocity shift and FWHM are fixed to the values obtained from the H$\beta$ complex fit; therefore, no uncertainties are reported for these quantities. The H$\alpha$ emission-line measurements are based on the \textit{NIRSpec} grating spectrum, while the H$\alpha$ absorption measurement is based on the \textit{NIRCam} grism spectrum, whose higher spectral resolution better reveals the absorption feature.

\end{minipage}
\end{table*}
\endgroup

\section*{METHODS}

\setcounter{figure}{0}
\renewcommand{\figurename}{Extended Data Figure}
\setcounter{table}{0}
\renewcommand{\tablename}{Extended Data Table}

\section*{Data and Observations} \label{sec:data}

The quasar JADES ID 209777, hereafter ``the Hatchling'', is located at $\alpha=03{:}32{:}38.03$ and $\delta=-27{:}46{:}26.56$. The object lies at a spectroscopic redshift of $z = 3.71$, measured from the Balmer line fitting. It is associated with the 7 Ms CDF-S X-ray source ID 715, which was classified as an AGN \cite{luo_chandra_2017}. This was targeted by NIRSpec as a potential AGN from the catalog of \cite{luo_chandra_2017}, where we refined the astrometry from the NIRCam imaging. The source was also identified as an AGN candidate in the MIRI SMILES survey \citep{lyu_active_2024} and subsequently observed in the JADES NIRSpec program \cite{deugenio_jades_2025,scholtz_jades_2025,curtis-lake_jades_2025}. This source was previously reported to have a redshift of $z = 3.711$ by \cite{maiolino_jwst_2024} and $z = 3.709$ by \cite{juodzbalis_jades_2025}. The slight difference between the reported redshifts may arise because the broad \OIIIwl{} component complicates the determination of the systemic redshift. In this section, we summarize the multiwavelength data used in this study.

\subsection*{MUSE Observations}

This field is covered by the MUSE Ultra Deep Field survey (ESO Programmes 094.A-0289(B), 095.A-0010(A), 096.A-0045(A,B), 1101.A-0127) \citep{bacon_muse_2023}, which provides some of the deepest ground-based integral field spectroscopy to date. The MUSE UDF-10 data reach a $3\sigma$ point-source line flux sensitivity of $\sim$ 1.5 $\times$ 10$^{-19}$ erg s$^{-1}$ cm$^{-2}$. We extracted a Ly$\alpha$ emission line map and 1D spectra for the Hatchling from the public AMUSED mosaic cubes. The source is clearly detected in Ly$\alpha$, with spatially extended emission out to $\sim$20 kpc and a significant velocity gradient across the halo. We also see tentative Ly$\beta$ emission, although at lower significance. 

\subsection*{Imaging: HST, NIRCam, and MIRI}

The Hatchling is located within the deep JADES imaging footprint and has extensive archival imaging from \textit{HST/ACS}, \textit{HST/WFC3} \cite{whitaker_hubble_2019}, JWST/NIRCam \cite{johnson_jwst_2026, robertson_jwst_2026, rieke_jades_2023}, and JWST/MIRI \cite{alberts_smiles_2024, rieke_smiles_2024}. The available \textit{HST} imaging includes F435W, F606W, F775W, F814W, F850LP, F105W, F125W, F140W, and F160W. The NIRCam imaging includes F090W, F115W, F150W, F182M, F200W, F210M, F277W, F335M, F356W, F410M, and F444W. The source is also detected in MIRI SMILES bands F560W, F770W, F1000W, F1280W, F1500W, F1800W, F2100W, and F2550W. 

The full UV-to-IR photometry is listed in Extended Data Table~\ref{tab:photometry}. These data provide robust constraints on the quasar's SED and host galaxy properties. 

\begin{table}[!ht]
\centering
\caption{Photometry of the Hatchling}
\label{tab:photometry}
\begin{tabular}{lcc}
\hline
Filter & Flux Density (nJy) & Uncertainty (nJy) \\
\hline
\textit{HST} F435W  & 15.6  & 3.6  \\
\textit{HST} F606W  & 34.3  & 2.8  \\
\textit{HST} F775W  & 79.0  & 3.7  \\
\textit{HST} F814W  & 93    & 14   \\
\textit{HST} F850LP & 174.4 & 7.6  \\
\textit{HST} F105W  & 271.5 & 6.8  \\
\textit{HST} F125W  & 623.2 & 7.9  \\
\textit{HST} F140W  & 936.8 & 7.9  \\
\textit{HST} F160W  & 1355  & 10   \\
\textit{JWST} F090W  & 151.2  & 7.9  \\
\textit{JWST} F115W  & 431.4  & 7.2  \\
\textit{JWST} F150W  & 1265.2 & 8.3  \\
\textit{JWST} F182M  & 2315.5 & 8.9  \\
\textit{JWST} F200W  & 2842.0 & 7.5  \\
\textit{JWST} F210M  & 3373.7 & 10.0 \\
\textit{JWST} F277W  & 5329.8 & 7.5  \\
\textit{JWST} F335M  & 5269.1 & 7.7  \\
\textit{JWST} F356W  & 5819.4 & 5.5  \\
\textit{JWST} F410M  & 7217.2 & 8.3  \\
\textit{JWST} F444W  & 7843.7 & 6.5  \\
\textit{JWST} F560W  & 11310  & 110  \\
\textit{JWST} F770W  & 21239  & 89   \\
\textit{JWST} F1000W & 35820  & 150  \\
\textit{JWST} F1280W & 52220  & 280  \\
\textit{JWST} F1500W & 64420  & 320  \\
\textit{JWST} F1800W & 77840  & 650  \\
\textit{JWST} F2100W & 84620  & 820  \\
\textit{JWST} F2550W & 97300  & 3800 \\
\hline
\end{tabular}
\end{table}

\subsection*{NIRSpec Spectroscopy}
The Hatchling was observed with JWST/NIRSpec as part of JADES Data Release~3 \citep{deugenio_jades_2025}. The spectroscopy covers \(0.6\)--\(5.3~\mu{\rm m}\), including both the low-resolution PRISM mode (\(R\sim100\)) and the three medium-resolution gratings G140M, G235M, and G395M (\(R\sim1000\)). The observations were conducted under PID 1180 on 2023-10-08, with 6215 s exposure time in each of the medium-resolution gratings and 7528 s in PRISM. 
The data were reduced using the JWST Calibration Pipeline v1.20.2 \citep{bushouse_jwst_2022} and CRDS reference files {\tt jwst\_1464.pmap}, with customized scripts for additional noise and artifact handling following the SMILES NIRSpec Data Release \citep{zhu_smiles_2026}.

The NIRSpec spectra reveal multiple rest-frame UV and optical emission lines, including broad H$\alpha$, H$\beta$, Mg II, [O III], and O I $\lambda 8446$, as well as absorption features in He I $\lambda 10830$, Na I D, and Mg II.


\subsection*{NIRCam Grism}
The Hatchling was observed with the NIRCam F277W grism under JWST GO program PID~7336 (PI: F.~Sun). The spectrum covers $2.4$ -- $3.2~\mu{\rm m}$, with an effective exposure time of 7,300~s. The data were reduced following \cite{sun_first_2023}\footnote{\url{https://github.com/fengwusun/nircam_grism}}. The grism spectrum has a spectral resolution of $R\sim1300$ near 3~$\mu$m, enabling the detection of an absorption feature associated with broad H$\alpha$ emission.

\subsection*{Other Data}
The Hatchling is detected in the 7~Ms \textit{Chandra} Deep Field-South data \citep{luo_chandra_2017} (ID 715). It is also detected in the public ASPECS ALMA Band~6 1.2~mm continuum image \citep{gonzalez-lopez_alma_2020, aravena_alma_2020, boogaard_alma_2020} (ID C08). In the radio, we use the ultradeep VLA imaging of the GOODS-S/HUDF region \citep{alberts_completing_2020, lyu_agn_2022}; the source detection is marginal at 6~GHz but clearly detected at 3~GHz. The combination of X-ray, millimeter, and radio emission further supports the AGN classification of this source.

\section*{Continuum Shape Fitting to Different Scenarios}
In this section, we explore the origins of the broad bump seen in the rest-frame UV--optical flux.

\subsection*{Dust attenuation}

\begin{figure*}[!ht]
    \centering
    \includegraphics[width=0.8\linewidth]{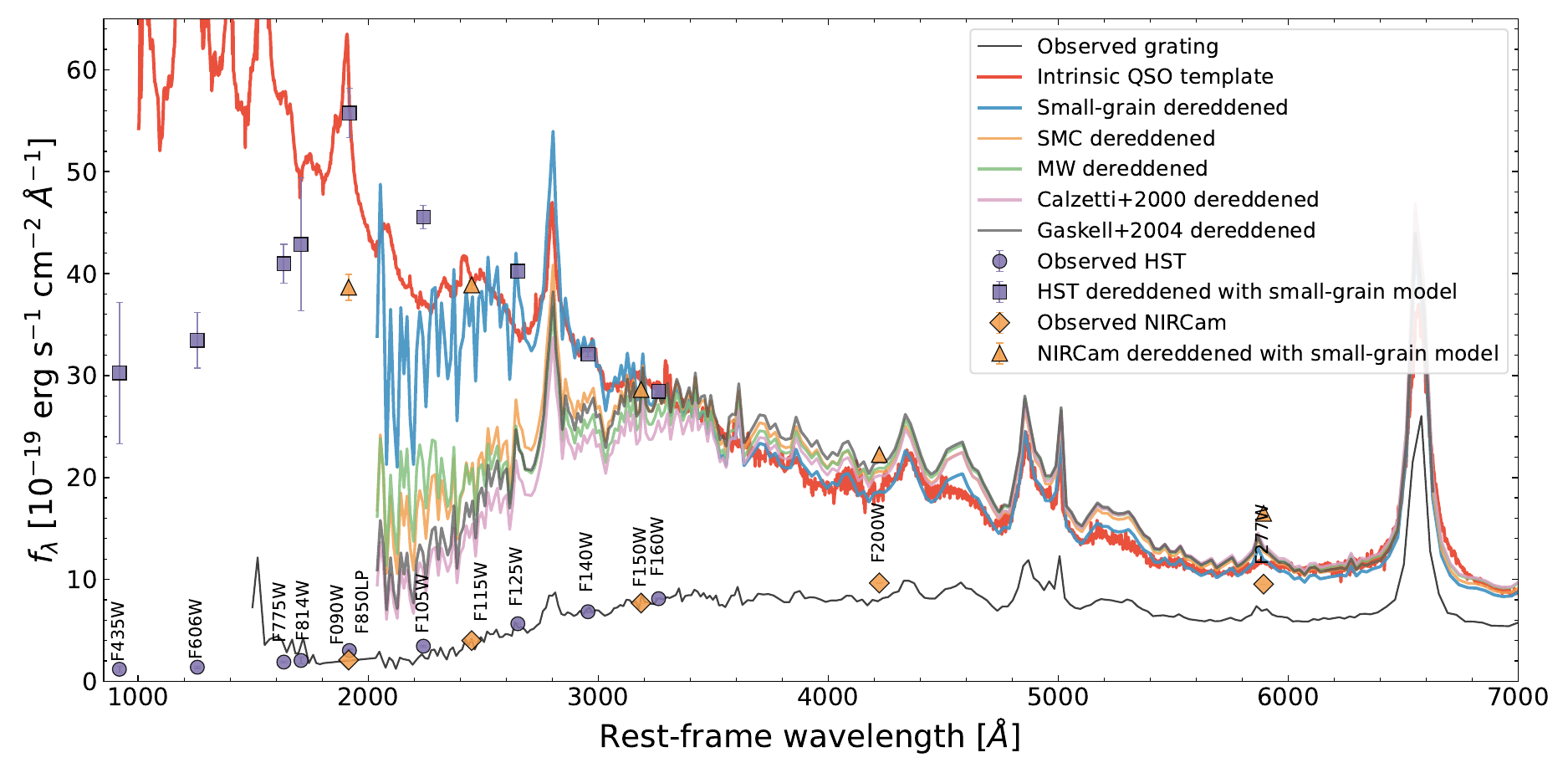}
    \caption{Comparison between the dereddened spectra and photometry obtained using different extinction/attenuation prescriptions and an intrinsic quasar composite. The black curve shows the observed rest-frame NIRSpec grating spectrum, while the red curve shows the intrinsic unobscured quasar composite \cite{selsing_x-shooter_2016}. The orange, green, pink, and gray curves show the spectra dereddened using the SMC extinction curve, the Milky Way extinction curve, the Calzetti attenuation law \cite{calzetti_dust_2000}, and the gray AGN extinction curve of Gaskell et al. \cite{gaskell_nuclear_2004}, respectively. The blue curve shows the spectrum dereddened using the best-fit small-grain model. The observed HST photometry and the corresponding dust-corrected photometry obtained with the best-fit small-grain model are shown as purple circles and squares, respectively. The observed NIRCam photometry and corresponding dust-corrected values are shown as gold diamonds and triangles, respectively. Standard extinction/attenuation laws do not recover the intrinsic quasar continuum shape over the full wavelength range, whereas the small-grain model provides a closer phenomenological match to the quasar composite.
    }
    \label{fig:dust_attenuation}
\end{figure*}

One possible interpretation is that the continuum is suppressed by dust reddening. We therefore tested whether the continuum of the Hatchling can be explained by dust attenuation of an intrinsic quasar spectrum. We first dereddened the observed spectrum using several commonly adopted extinction or attenuation laws, including the Milky Way (MW), Small Magellanic Cloud (SMC), attenuation laws from Calzetti et al. \cite{calzetti_dust_2000} and AGN nuclear reddening curve from \cite{gaskell_nuclear_2004}. As shown in Extended Data Figure~\ref{fig:dust_attenuation}, these standard laws do not recover an intrinsic SED consistent with the unreddened quasar composite spectrum.

Steep far-ultraviolet extinction curves produced by abundant small dust grains have been proposed as a possible explanation for the SEDs of red quasars \citep{jiang_anomalously_2013, sun_jades_2026,li_steep-extinction_2026}. We therefore also considered a small-grain extinction model following the procedure of \cite{li_steep-extinction_2026}. In this model, the extinction curve, normalized by $A_V$, is parameterized as a sum of Drude profiles:
\begin{multline}
A_{\lambda}/A_V =
\frac{c_1}
{(\lambda/0.08)^{c_2} + (0.08/\lambda)^{c_2} + c_3}
\\
+ \frac{
233\left[1 - c_1/(6.88^{c_2} + 0.145^{c_2} + c_3) - c_4/4.60\right]}
{(\lambda/0.046)^2 + (0.046/\lambda)^2 + 90}
\\
+ \frac{c_4}
{(\lambda/0.2175)^2 + (0.2175/\lambda)^2 - 1.95}.
\label{eq:small_grain_extinction}
\end{multline}
where $\lambda$ is in units of $\mu$m \cite{li_dust_2008}.

We adopted the intrinsic quasar composite spectrum of \cite{selsing_x-shooter_2016}. To account for object-to-object variations in the intrinsic continuum slope, we allowed an additional power-law tilt:
\begin{equation}
f_{\lambda,{\rm int}} =
N_{\rm norm} f_{\lambda,{\rm template}}
\left(\frac{\lambda}{\lambda_0}\right)^{\alpha_0-\alpha_{\lambda}},
\label{eq:qso_template_tilt}
\end{equation}
where $N_{\rm norm}$ is a normalization factor, $\lambda_0$ is the reference wavelength, and $\alpha_{\lambda}$ controls the continuum slope relative to the template slope $\alpha_0$.

The attenuated model spectrum is then
\begin{equation}
f_{\lambda,{\rm obs}} =
f_{\lambda,{\rm int}}\,10^{-0.4 A_{\lambda}}.
\label{eq:attenuated_model}
\end{equation}
We fitted the NIRSpec grating spectrum with this attenuated quasar-template model and derived the best-fit small-grain model.

We then applied the best-fitting extinction curve to deredden the grating spectrum, as well as the HST and NIRCam photometry, and compared the inferred intrinsic SED with the unreddened quasar template. As shown in Extended Data Figure~\ref{fig:dust_attenuation},  when the grating spectrum is dereddened using a small-grain extinction model, the resulting spectrum closely matches the intrinsic QSO template.

The continuum fit yields an effective attenuation of $A_{V,\mathrm{cont}}=0.74$. Using the narrow-line Balmer decrement, assuming an intrinsic Case~B ratio of $(\ha/\hb)_0=2.86$ and the SMC extinction curve \cite{gordon_quantitative_2003}, we infer a substantially larger nebular attenuation of $A_{V,\mathrm{NLR}}=1.81^{+1.25}_{-0.89}$. This value is broadly consistent, within the uncertainties, with $A_{V,\mathrm{NLR}}=2.56^{+0.83}_{-0.70}$ reported by \cite{juodzbalis_jades_2025}. The difference between $A_{V,\mathrm{cont}}$ and $A_{V,\mathrm{NLR}}$ may reflect different dust columns and sightlines toward the compact continuum-emitting region and the more extended narrow-line gas, together with a patchy dust geometry and the different attenuation prescriptions adopted in the two measurements.

\begin{figure*}[!ht]
    \centering
    \includegraphics[width=0.9\linewidth]{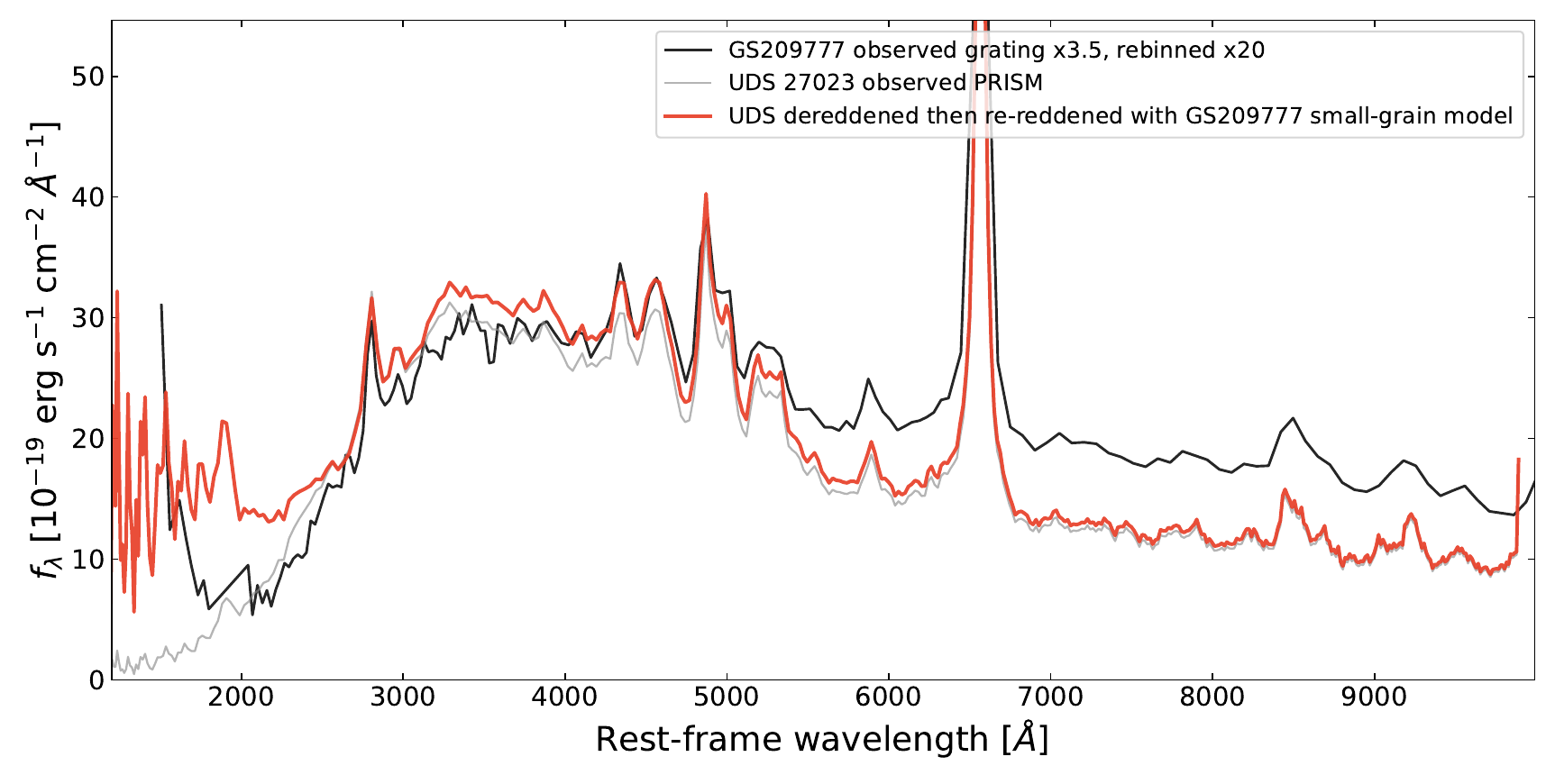}
    \caption{Comparison of the NIRSpec/PRISM spectrum of UDS 27023 with the NIRSpec/grating spectrum of GS209777, the Hatchling. The gray curve shows the observed PRISM spectrum of UDS 27023. The red curve shows the same spectrum after dereddening with the best-fit dust model for UDS 27023 and then re-reddening with the dust model derived for the Hatchling. The black curve shows the grating spectrum of the Hatchling, scaled by a factor of 3.5 and rebinned over 20 pixels for display. This cross-reddening exercise shows that UDS 27023 can reproduce the overall continuum shape of the Hatchling over rest-frame $2500$--$5000$ \AA. In contrast, the Hatchling shows additional emission relative to the transformed UDS 27023 spectrum at wavelengths below $\sim2500$ \AA\ and above $\sim5000$ \AA.}    
    \label{fig:cross_reddening}
\end{figure*}

UDS 27023 is another quasar with an unusually steep extinction curve, at $z = 4.556$, whose reddening has been attributed to small dust grains \cite{li_steep-extinction_2026}. Its SED shape is similar to that of the Hatchling. To test whether the two sources could be described by similar attenuation physics, we performed a cross-reddening experiment. We first dereddened the observed UDS 27023 PRISM spectrum using its best-fitting small-grain extinction model to infer an intrinsic spectrum. We then reddened this inferred intrinsic spectrum using the best-fitting small-grain extinction model derived for the Hatchling. The resulting re-reddened spectrum was compared directly with the observed grating spectrum of the Hatchling in Extended Data Figure~\ref{fig:cross_reddening}.

This comparison shows that the transformed UDS 27023 spectrum reproduces the continuum shape of the Hatchling only over the limited rest-frame range of $2500$--$5000$ \AA. The agreement does not extend to shorter or longer wavelengths, where the Hatchling shows excess emission relative to the transformed UDS 27023 spectrum. Thus, differences in dust attenuation may explain part of the observed SED similarity, but the match is restricted and does not by itself imply that the two objects belong to the same physical class.

\begin{figure}
    \centering
    \includegraphics[width=0.9\linewidth]{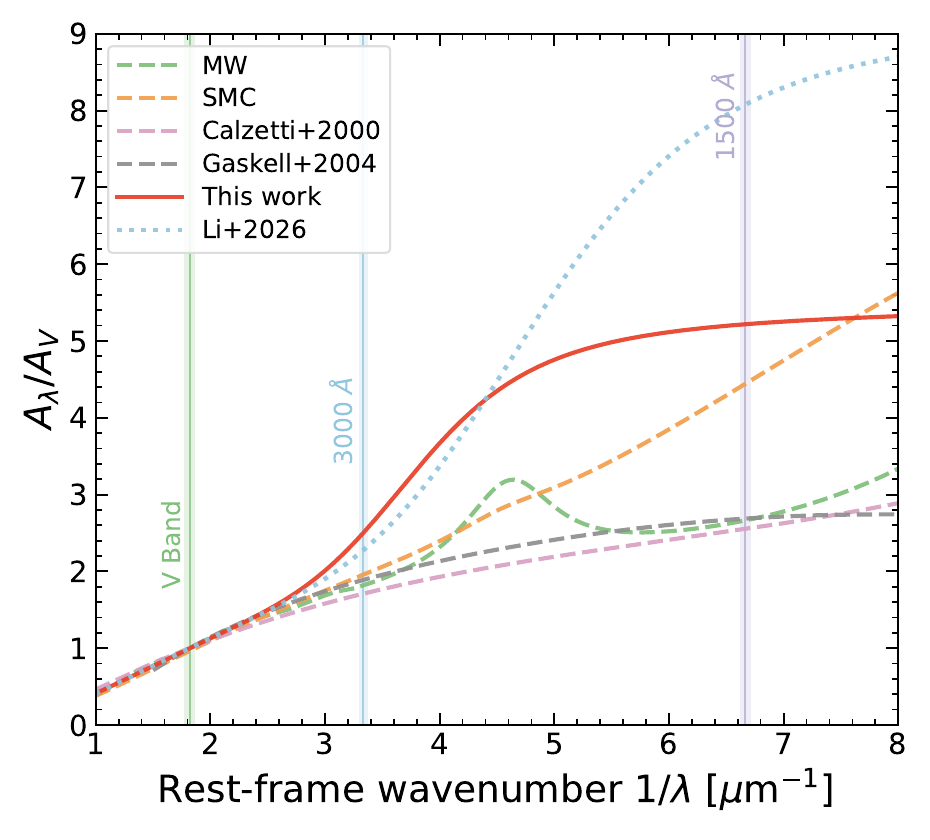}
    \caption{
    Comparison of the dust curve derived in this work with commonly used extinction and attenuation curves. The curves show \(A_\lambda/A_V\) as a function of rest-frame wavenumber \(1/\lambda\). The MW and SMC extinction curves, the Calzetti et al. \citep{calzetti_dust_2000} attenuation law, and the AGN nuclear reddening curve from \cite{gaskell_nuclear_2004} are shown as the purple, orange, pink, and gray dashed lines, respectively. The red solid curve shows the dust curve inferred in this work, while the blue dotted curve shows the small-grain dust model from \cite{li_steep-extinction_2026}. Vertical lines mark the \(V\) band, 3000~\AA, and 1500~\AA. This comparison shows that the reddening required for the Hatchling is stronger in the UV than standard extinction/attenuation laws, but does not require the extremely steep UV rise inferred for UDS 27023.}
    \label{fig:extinction_curves}
\end{figure}

Extended Data Figure~\ref{fig:extinction_curves} places the extinction curve used for the Hatchling in context by comparing it with the dust model adopted for UDS 27023 by \cite{li_steep-extinction_2026} and with several commonly used prescriptions. These include the SMC and Milky Way extinction curves, the Calzetti et al. \cite{calzetti_dust_2000} attenuation law, and the AGN nuclear reddening curve from \cite{gaskell_nuclear_2004}. If the red continuum of our source is interpreted as dust reddening, it requires enhanced extinction toward shorter wavelengths relative to standard laws, but the inferred curve is not as steep as that applied to UDS 27023.

We estimate the dust mass from the ALMA Band~6 1.2~mm continuum flux density \cite{boogaard_alma_2020, gonzalez-lopez_alma_2020, aravena_alma_2020} assuming optically thin modified-blackbody emission:
\begin{equation}
M_{\rm d} =
\frac{S_{\nu_{\rm obs}} D_L^2}
{(1+z)\,\kappa_{850,\mu{\rm m}}
\left(\nu_{\rm rest}/\nu_{850,\mu{\rm m}}\right)^{1.8}
B_{\nu_{\rm rest}}(T_{\rm d})}.
\end{equation}
Here \(S_{\nu_{\rm obs}}\) is the observed 1.2~mm flux density, \(D_L\) is the luminosity distance, \(\nu_{\rm rest}=(1+z)\nu_{\rm obs}\), and \(B_{\nu_{\rm rest}}(T_{\rm d})\) is the Planck function. We adopt a dust mass absorption coefficient of \(\kappa_{850\,\mu{\rm m}}=0.077~{\rm m^2~kg^{-1}}\) \citep{draine_optical_1984} and fix the dust emissivity index to \(\beta=1.8\). Allowing \(T_{\rm d}=25-55~{\rm K}\), the measured \(S_{1.2\,\mathrm{mm}}=163\pm10~\mu{\rm Jy}\) implies \(M_{\rm d}\approx(2.1-9.9)\times10^7~M_\odot\), with \(M_{\rm d}\approx4.6\times10^7~M_\odot\) for \(T_{\rm d}=35~{\rm K}\). SED fitting provided a total infrared luminosity of
$\log_{10}(L_{\rm IR}/L_\odot)=12.53\pm0.16$ and
$\log_{10}({\rm SFR}/M_\odot~{\rm yr^{-1}})=2.48\pm0.16$ \cite{boogaard_alma_2020, gonzalez-lopez_alma_2020, aravena_alma_2020},
corresponding to
${\rm SFR}\approx300^{+136}_{-93}~M_\odot~{\rm yr^{-1}}$.
The high infrared luminosity places the source in the ULIRG regime.

An additional constraint comes from the single-epoch black hole (BH) mass estimates (see \textit{Spectral line fitting and black hole mass} below). The observed masses inferred from \MgII{}, \hb{}, and \ha{} are broadly consistent within the systematic uncertainty of virial mass calibrations, with $\log(M_{\rm BH}/M_\odot)=8.63\pm0.14$, $8.84\pm0.09$, and $8.38\pm0.02$, respectively. If the continuum were strongly dust attenuated, the intrinsic line-continuum luminosities would be higher, increasing the corresponding virial BH masses as $\Delta\log M_{\rm BH}=0.2A_\lambda$, assuming that the line widths are unchanged. For the small-grain attenuation model, the inferred corrections are $\Delta\log M_{\rm BH}=0.458$, $0.207$, and $0.161$ dex for \MgII{}, \hb{}, and \ha{}, respectively. The corrected masses would therefore become $\log(M_{\rm BH}/M_\odot)=9.09\pm0.14$, $9.05\pm0.09$, and $8.51\pm0.02$. The dust-corrected values therefore increase the tension among the Mg\,{\sc ii}, H\(\beta\), and H\(\alpha\) mass estimates, although the discrepancy remains comparable to the systematic uncertainty of single-epoch virial BH mass calibrations.

\subsection*{Empirical modified-blackbody continuum fit}
\begin{figure}[!ht]
    \centering
    \includegraphics[width=0.9\linewidth]{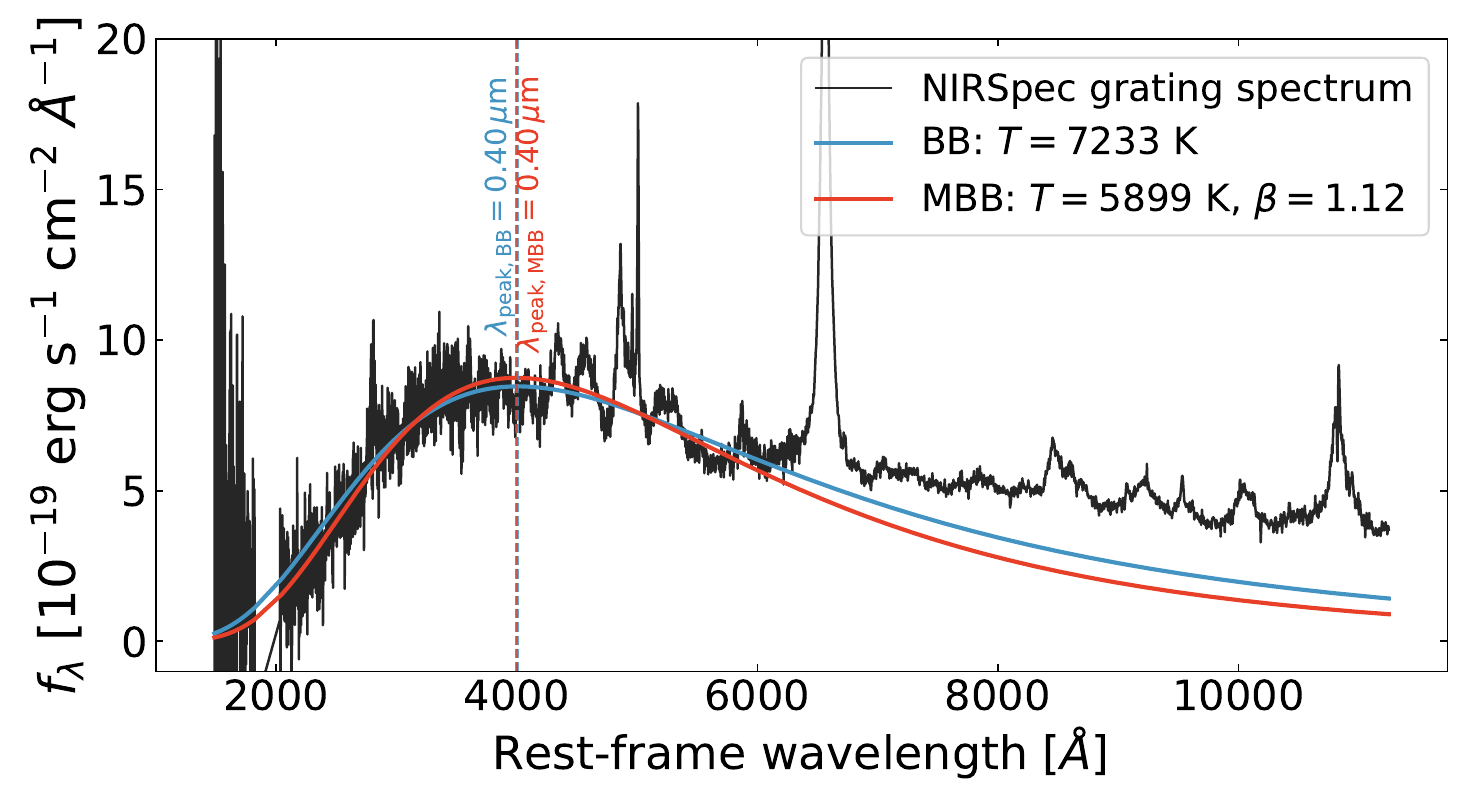}
    \caption{Rest-frame NIRSpec grating spectrum of the Hatchling and its thermal-continuum fits. The black curve shows the stitched grating spectrum, the red curve shows the best-fitting modified blackbody (MBB) continuum with $T=5899$~K and $\beta=1.12$, and the blue curve shows the best-fitting blackbody (BB) model with $T=7233$~K. The red and blue dashed lines mark the peak wavelengths of the MBB and BB continua, respectively. The broad rest-frame UV-to-optical bump can be empirically described by a modified blackbody curve. }
    \label{fig:mbbfit}
\end{figure}

Given the presence of \ha{} and \HeI{} absorption, we also considered whether dense gas close to the nucleus could contribute to the continuum shape through thermal-like reprocessing, although this interpretation is not uniquely required by the data. We fitted the NIRSpec grating spectrum with a modified blackbody model \cite{graaff_little_2025}, as shown in Extended Data Figure~\ref{fig:mbbfit}. 

The fit was performed in the rest frame over $2100$--$6000$\,\AA. Since the spectrum is measured in $f_\lambda$, we used
\begin{equation}
f_\lambda(\lambda)
=
A_\lambda
\frac{B_\lambda(\lambda,T_{\rm bb})}
     {B_\lambda(\lambda_{\rm norm},T_{\rm bb})}
\left(\frac{\lambda}{\lambda_{\rm norm}}\right)^{-\beta},
\end{equation}
where $B_\lambda$ is the Planck function, $T_{\rm bb}$ is the effective temperature, and $\beta$ modifies the slope of the power law. We set the normalization wavelength to $\lambda_{\rm norm}=4000\,$\AA, so that $A_\lambda$ corresponds to the model flux density at $\lambda_{\rm norm}$. We excluded wavelength intervals affected by strong emission-line complexes, including \MgII{}, \hb{}, \OIII{} and \HeIop{}.

The best-fitting model gives
$T_{\rm bb}=5899^{+43}_{-42}\,{\rm K}$ and
$\beta =1.12\pm0.04$.
The peak wavelength was measured from the maximum of the best-fitting $f_\lambda$ modified-blackbody model, giving
$\lambda_{\rm peak} = 0.40\,\mu{\rm m}$ in the rest frame. We computed the modified-blackbody luminosity by integrating the best-fitting model over rest-frame $0.1$--$3\,\mu{\rm m}$, obtaining
$\log (L_{\rm bb}/{\rm erg\,s^{-1}}) \simeq 44.77$.

We also fit the continuum with a simple blackbody model, obtaining 
$T_{\rm bb}=7233\pm9~{\rm K}$. However, the modified blackbody model provides a better fit, with a much lower Bayesian information criterion (BIC), with 
$\Delta{\rm BIC}={\rm BIC}_{\rm MBB}-{\rm BIC}_{\rm BB}=-728$. 
We therefore use the modified blackbody model  as an empirical description of the continuum curvature.

\begin{figure*}
    \centering
    \includegraphics[width=0.8\linewidth]{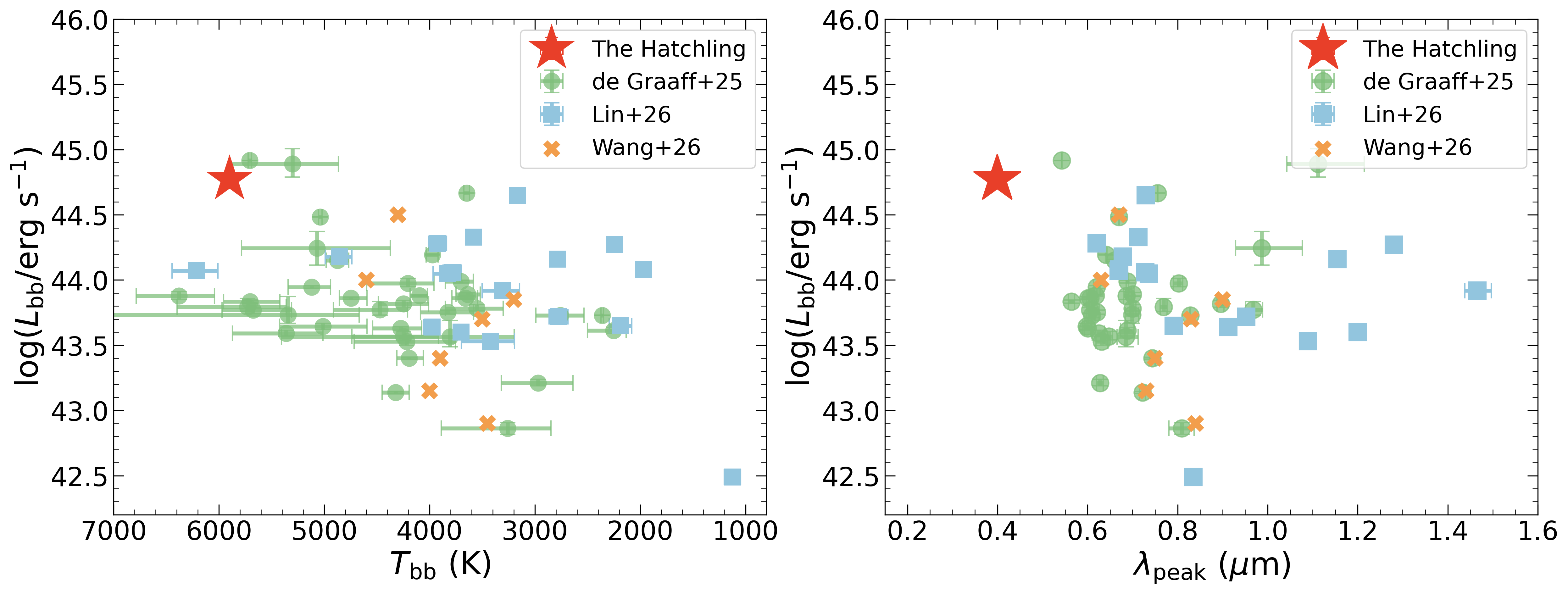}
    \caption{H--R diagrams of the Hatchling compared with LRD samples. \textbf{Left:} H--R diagram showing the total blackbody luminosity, $L_{\rm bb}$, as a function of effective temperature, $T_{\rm bb}$. The Hatchling is compared with LRD samples from \cite{graaff_little_2025} (green plus signs), \cite{lin_lrds2_2026} (blue squares), and \cite{wang_water_2026} (orange crosses).
    \textbf{Right:} H--R diagram showing $L_{\rm bb}$ as a function of peak wavelength, $\lambda_{\rm peak}$. The comparison samples and symbols are the same as in the left panel.
    The fitted red continuum of the Hatchling has a luminosity comparable to compact red AGNs but a warmer effective temperature and shorter peak wavelength.}
    \label{fig:hr_diagram}
\end{figure*}

Extended Data Figure~\ref{fig:hr_diagram} compares the Hatchling with LRD samples from \cite{graaff_little_2025, lin_lrds2_2026, wang_water_2026} in the modified-blackbody H--R diagram. The bulk of the LRD population peaks at $\lambda_{\rm peak}$\,$\sim$\,0.6--1.0~$\mu$m, corresponding to fitted modified-blackbody temperatures of order $T_{\rm bb}$\,$\sim5000$~K. The Hatchling instead peaks at a shorter wavelength in this empirical parameterization, corresponding to a higher fitted effective temperature than typical LRDs.

\subsection*{BAL-like absorption}
Several trough-like features are present in the rest-frame UV spectrum and may resemble those seen in mini-BAL quasars. However, the evidence is not sufficient for a robust BAL or mini-BAL classification. The putative troughs are relatively weak and are affected by limited signal-to-noise, continuum-placement uncertainty, and possible contamination from Fe\,{\sc ii} emission, which produces a structured pseudo-continuum around the Mg\,{\sc ii} region. We therefore regard the mini-BAL interpretation as possible but not conclusive, and do not use it as a primary component of our physical interpretation.

\section*{Spectral line fitting and black hole mass measurements}
\label{sec:line_fitting}
We first modelled the quasar pseudo-continuum using the public code \textsc{PyQSOFit} \citep{shen_sloan_2019,guo_pyqsofit_2018}. The model includes a power-law continuum and empirical Fe~\textsc{ii} templates, and was fit to the observed-frame spectrum over $13000$--$47000$~\AA{}~using continuum windows chosen to avoid strong emission-line features. We subtracted the best-fitting pseudo-continuum and adopted the systemic redshift $z$\,$=$\,$3.711$ inferred by Balmer line fitting.

We then modelled the emission and absorption features with Gaussian components using an MCMC approach. For the \hb+\OIII\ complex, we fit the pseudo-continuum-subtracted spectrum, because the optical Fe~\textsc{ii} emission contributes significantly in this wavelength range. The \OIII\ doublet was modelled either with a single kinematic component or with narrow and broad kinematic components, with the \OIII\ $\lambda5008/\lambda4960$ flux ratio fixed to 2.98. The \hb\ line was modelled with one narrow component and up to two broad components. The velocity shift and FWHM of the narrow \hb\ component were tied to those of the narrow \OIII\ component. We required narrow components to have $\mathrm{FWHM}<500~\mathrm{km~s^{-1}}$ and broad components to have $\mathrm{FWHM}>500~\mathrm{km~s^{-1}}$. We compared the alternative models using the BIC, and adopted the model with the lowest BIC. The preferred model consists of narrow and broad \OIII\ components and one narrow plus two broad \hb\ components.

For the other line complexes, we fit the original NIRSpec grating spectrum directly and included a local power-law continuum in each fitting window. The continuum parameters were allowed to vary independently for each line complex. This approach was used for \MgII{}, \HeIop{}, \ha{}, \OI{}, \Pagamma{}, and \HeI{}. We did not include an explicit UV Fe~\textsc{ii} template in the \MgII{} fitting region; instead, we modelled the local continuum with a power law. Since Fe~\textsc{ii} emission can contribute beneath \MgII{}, the \MgII{} measurements should be regarded as empirical.

We also modelled the \ha{} complex separately with one narrow component and up to two broad components. The velocity shift and FWHM of the narrow \ha{} component were fixed to the narrow-component kinematics inferred from the \hb{}+\OIII{} fit. The \NII{} doublet was modelled with the same narrow-component kinematics as \ha{}, and the \NII{} $\lambda6585/\lambda6549$ flux ratio was fixed to 2.96. The preferred \ha{} model consists of one narrow component, two broad components, and narrow \NII{} emission.

The source also has NIRCam WFSS obtained with the F277W grism. The grism spectrum covers most of the \ha{} complex, although the red side of the line is partially outside the wavelength coverage. We fit the grism spectrum independently using a local power-law continuum, one narrow \ha{} component, and up to two broad \ha{} components. The velocity shift and FWHM of the narrow \ha{} component were fixed to the narrow-component values inferred from the \hb{}+\OIII{} fit. Because the grism spectrum shows an absorption feature superposed on the \ha{} emission profile near the line peak, we included an additional Gaussian absorption component.

For the remaining isolated emission lines, including \HeIop{}, \OI{}, \HeI{}, and \Pagamma{}, we modelled the local continuum with a power law and fit each line with up to two Gaussian components. Absorption features associated with \MgII\ and \nad\ were modelled empirically using Gaussian absorption components on top of the local emission-line baseline. We report the flux, velocity shift, FWHM, and rest-frame equivalent width of each fitted component in Table~\ref{tab:line_measurements}.

Adopting the velocity width measured for \OI{}, $\rm{FWHM} = 3534\,km\,s^{-1}$, we do not detect the corresponding O\,$\textsc{i}\,\lambda1304$ feature. We derive a 3 $\sigma$ upper limit of $F_{\rm{O}\,\textsc{I}\,\lambda1304}$~$<$~9.66\,$\times$\,$10^{-19}$\,erg\,$\rm{s}^{-1}$\,$\rm{cm}^{-2}$. Compared with the 8.76\,$\times$\,$10^{-17}$\,erg\,$\rm{s}^{-1}$\,$\rm{cm}^{-2}$ expected for an O\,$\textsc{I}$ $(\lambda1304/\lambda8446)$ photon ratio of unity, this implies $N_{1304}/N_{8446} < 0.012$. The suppressed UV O I emission is consistent with significant attenuation toward the O I-emitting region, although resonant-transfer effects and non-fluorescent contributions to \OI{} may also affect the ratio.

We estimate the black hole mass independently from the broad \ha{}, \hb{}, and \MgII{} emission-line fits. For \ha{}, we used the virial calibration from \cite{reines_dwarf_2013}:
\begin{equation}
\begin{aligned}
\log_{10}\left(\frac{M_{\rm BH,H\alpha}}{M_\odot}\right)
&= 6.57 + \log_{10}(\epsilon) \\
&\quad + 0.47 \log_{10}\left[
\frac{L_{{\rm H}\alpha,{\rm broad}}}
{10^{42}\ {\rm erg\ s^{-1}}}
\right] \\
&\quad + 2.06 \log_{10}\left[
\frac{{\rm FWHM}_{{\rm H}\alpha,{\rm broad}}}
{1000\ {\rm km\ s^{-1}}}
\right],
\end{aligned}
\end{equation}
where $\epsilon$ is the scale factor, for which we adopt $\epsilon=1.075$, consistent with \cite{reines_relations_2015}.

For \hb{}, we used the single-epoch virial calibration from \cite{vestergaard_determining_2006}:
\begin{equation}
\begin{aligned}
\log_{10}\left(\frac{M_{\rm BH, H\beta}}{M_\odot}\right)
&= 6.91 + 0.50 \log_{10}\left[
\frac{\lambda L_{\lambda}(5100\,\text{\AA})}
{10^{44}\ {\rm erg\ s^{-1}}}
\right] \\
&\quad + 2.00 \log_{10}\left[
\frac{{\rm FWHM}_{\rm H\beta, broad}}
{1000\ {\rm km\ s^{-1}}}
\right].
\end{aligned}
\end{equation}

For \MgII{}, we used the single-epoch virial calibration from \cite{vestergaard_mass_2009}:
\begin{equation}
\begin{aligned}
\log_{10}\left(\frac{M_{\rm BH, Mg\,II}}{M_\odot}\right)
&= 6.86 + 0.50 \log_{10}\left[
\frac{\lambda L_{\lambda}(3000\,\text{\AA})}
{10^{44}\ {\rm erg\ s^{-1}}}
\right] \\
&\quad + 2.00 \log_{10}\left[
\frac{{\rm FWHM}_{\rm Mg\,II, broad}}
{1000\ {\rm km\ s^{-1}}}
\right].
\end{aligned}
\end{equation}

For line profiles modelled with multiple broad Gaussian components, we measured an effective FWHM from the summed broad-line profile rather than from any individual Gaussian component. Specifically, for each MCMC posterior draw, we
constructed the total broad-line profile by summing all broad Gaussian components, measured the wavelengths at which this summed profile reaches half of its maximum value, and converted the corresponding wavelength width to a
velocity FWHM. For \ha{}, we integrated the summed broad-line model to obtain the broad-line flux and converted it to luminosity using the luminosity distance. For \hb{} and \MgII{}, we used the continuum luminosities at rest-frame 5100 and 3000~\AA{}, respectively.

For \ha{}, we adopted the fit to the NIRSpec spectrum rather than the NIRCam/WFSS grism spectrum because the grism spectrum does not fully cover the red side of the \ha\ emission profile. The black hole masses inferred from \ha, \hb, and \MgII\ are $\log(M_{\rm BH}/M_\odot)=8.38 \pm 0.02$, $8.84 \pm 0.09$, and $8.63 \pm 0.14$, respectively. For all three estimates, uncertainties were propagated by applying the above procedure to each posterior draw of the line model and reporting the median and 16th--84th percentile range of the resulting BH mass distribution. These quoted uncertainties reflect only the statistical uncertainties from the spectral modeling and do not include the systematic uncertainty of single-epoch virial black hole mass calibrations, which is typically $\sim0.46$ dex \cite{park_lick_2012}.

Previous analyses of this source have reported \ha{}-based masses of $\log(M_{\rm BH}/M_\odot)=8.90\pm0.30$ \citep{juodzbalis_jades_2025} and $8.23\pm0.11$ \citep{hoshi_evolutionary_2025}. The spread among the published values likely arises from differences in spectral extraction, line-profile decomposition, the treatment of the extended \ha{} wings and outflow contribution, and the adopted virial calibration. We take our NIRSpec-based \ha{} estimate for internal consistency, while treating the range among the available measurements as an indication of the additional systematic uncertainty associated with modelling the complex \ha{} profile.

\section*{Outflow masses and mass-outflow-rate estimates}

For \OIII, we estimate the characteristic velocity of the outflowing gas from the broad \OIII\ component following \cite{carniani_jades_2024}:
\begin{equation}
v_{\rm out}=|v_{\rm broad}-v_{\rm narrow}|+2\sigma_{\rm broad},
\end{equation}
where $v_{\rm broad}$ and $v_{\rm narrow}$ are the best-fitting velocity shifts of the broad and narrow \OIII\ components, respectively, and $\sigma_{\rm broad}={\rm FWHM}_{\rm broad}/2.355$.

We estimate the ionized-gas outflow mass and mass outflow rate following \cite{liu_frequent_2025}. The ionized outflow mass is given by
\begin{equation}
M_{\rm out}=5.3\times10^{8}\,C_e\,
\frac{L_{44}({\rm [O\,III]})}{n_{e,2}\,10^{[{\rm O/H}]}}\ M_\odot,
\end{equation}
where $L_{44}({\rm [O\,III]})$ is the luminosity of the broad [O~\textsc{iii}] $\lambda5008$ component in units of $10^{44}~{\rm erg~s^{-1}}$, $n_{e,2}=n_e/(100~{\rm cm^{-3}})$, and $[{\rm O/H}]$ is the oxygen abundance relative to solar. The parameter $C_e\equiv\langle n_e\rangle^2/\langle n_e^2\rangle$ is the electron-density clumping factor. We assume $C_e=1$, $n_e=200~{\rm cm^{-3}}$ corresponding to $n_{e,2}=2$, and solar metallicity, $[{\rm O/H}]=0$.

The mass outflow rate is then calculated as
\begin{equation}
\dot{M}_{\rm out}=\frac{M_{\rm out}v_{\rm out}}{R_{\rm out}},
\end{equation}
where $R_{\rm out}$ is the characteristic outflow radius. We adopt $R_{\rm out}=1~{\rm kpc}$. With these assumptions, we obtain $M_{\rm out}=3.0^{+0.2}_{-0.2}\times10^{6}~M_\odot$, $v_{\rm out}=2100^{+220}_{-200}~{\rm km~s^{-1}}$, and $\dot{M}_{\rm out}=6.4^{+0.8}_{-0.7}$ \Msunyr.

We also estimate the mass outflow rates of the lower-ionization gas traced by \MgII\ absorption and the neutral gas traced by \nad\ absorption following \cite{davies_jwst_2024}:
\begin{equation}
\begin{aligned}
\dot{M}_{\rm out}
=&\ 11.45
\left(\frac{C_{\Omega}C_f}{0.4}\right)
\left(\frac{N_{\rm H}}{10^{21}\ {\rm cm}^{-2}}\right) \\
&\times
\left(\frac{r_{\rm out}}{1\ {\rm kpc}}\right)
\left(\frac{v_{\rm out}}{200\ {\rm km\ s}^{-1}}\right) \mathrm{M}_{\odot}\  \mathrm{yr}^{-1}.
\end{aligned}
\end{equation}
For consistency with the \OIII\ analysis, we define the absorption-line outflow velocity as
\begin{equation}
v_{\rm out}=|v_{\rm abs}|+2\sigma_{\rm abs},
\end{equation}
where $v_{\rm abs}$ is the velocity shift of the absorption component relative to the systemic redshift, and $\sigma_{\rm abs}={\rm FWHM}_{\rm abs}/2.355$ is the corresponding velocity dispersion.

Since our empirical Gaussian absorption models do not directly constrain the gas column density, we assume a fiducial hydrogen column density of $N_\mathrm{H}=10^{21}\ \mathrm{cm}^{-2}$ for both \MgII\ and \nad. We adopt $C_\Omega=0.5$, corresponding to an outflow covering 50\% of the full solid angle, and $C_f=0.3$, the median covering factor inferred for AGN-driven neutral-gas outflows in $z\sim2$ galaxies \citep{davies_jwst_2024}. We assume an outflow radius of $r=1\ \mathrm{kpc}$, consistent with the value adopted for the \OIII\ outflow. Under these assumptions, we derive a mass outflow rate $\dot{M}_\mathrm{out}$ = $15.3^{+4.8}_{-4.3}$ \Msunyr\ for \MgII\ and $\dot{M}_\mathrm{out}$ = $15.1^{+1.0}_{-0.8}$ \Msunyr\ for \nad{}.

The quoted uncertainties only propagate the statistical uncertainties from the line-profile fits and do not include systematic uncertainties associated with the assumed $N_\mathrm{H}$, $C_\Omega$, $C_f$, or $r_{\rm out}$. These values should therefore be interpreted as order-of-magnitude estimates, particularly for the absorption-line phases where the gas column density and geometry are not directly constrained.

\section*{Comparison of the Ly$\alpha$ halo}

\begin{table}
\centering
\caption{\Lya\ and UV continuum measurements from MUSE.}
\label{tab:lya}
\begin{tabular}{lcc}
\hline
Parameter & Value & Notes \\
\hline
$L_{\rm Ly\alpha}$ 
& $(2.0 \pm 0.3)\times10^{42}\ {\rm erg\ s^{-1}}$ 
& $a$ \\

${\rm EW}_0({\rm Ly}\alpha)$ 
& $36.8 \pm 16.1\ \text{\AA}$ 
& $b$ \\

${\rm FWHM}_{\rm Ly\alpha}$ 
& $670^{+40}_{-70}\ {\rm km\ s^{-1}}$ 
& $c$ \\

$\Delta v_{\rm peak}$ 
& $25^{+65}_{-0}\ {\rm km\ s^{-1}}$ 
& $d$ \\

$M_{\rm UV,1500}$ 
& $-19.0 \pm 0.2\ {\rm mag}$ 
& $e$ \\
\hline
\end{tabular}

\tablecomments{
$^{a}$ \Lya\ luminosity measured from the continuum-subtracted MUSE cube using a $3\arcsec$-radius aperture.
$^{b}$ Rest-frame \Lya\ equivalent width (\AA{}). The \Lya\ flux is measured from the continuum-subtracted cube, while the continuum flux density is measured from the original MUSE cube using the same $3\arcsec$ aperture.
$^{c}$ FWHM of the \Lya\ line measured from the $3\arcsec$ aperture spectrum extracted from the continuum-subtracted cube.
$^{d}$ Velocity offset of the \Lya\ peak relative to the systemic redshift.
$^{e}$ Rest-frame UV absolute magnitude at 1500 \AA{}, measured from the original MUSE cube using a $1\arcsec$-radius aperture.
}
\end{table}

We used the MUSE datacube to extract an $8\arcsec \times 8\arcsec$ cutout centered on the target. Two nearby lower-redshift sources lie close to the extended \Lya\ emission and contaminate the cutout with continuum light. We therefore estimated the continuum in each spaxel by applying a median filter along the spectral axis with a window of $\pm150$ spectral pixels, and subtracted the filtered cube from the original datacube to construct a continuum-subtracted emission-line cube, following the method described in \cite{wisotzki_extended_2016}. 
\begin{figure*}
    \centering
    \includegraphics[width=0.8\linewidth]{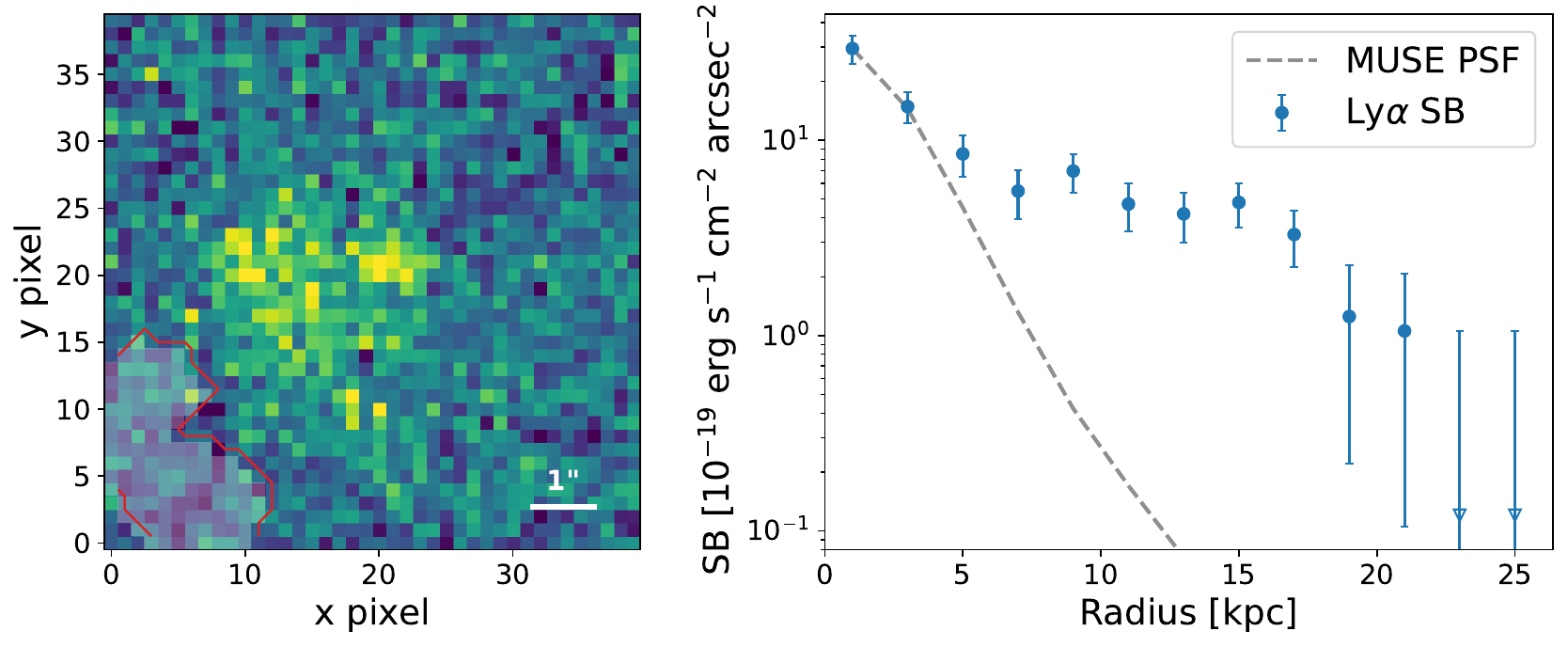}
    \caption{
    \textbf{Left:} Narrowband (NB) image of \Lya\ emission. The red shaded regions indicate areas masked to exclude contaminating sources.
    \textbf{Right:} Radial surface-brightness profile of the \Lya\ emission. Blue points show the azimuthally averaged \Lya\ surface brightness measured in concentric circular annuli, with $1\sigma$ uncertainties; triangles mark annuli with negative measured values. The gray curve shows the radial profile of the MUSE PSF. The \Lya\ surface-brightness profile is more extended than the PSF, indicating that the \Lya\ halo extends to a projected radius of $\sim20$ kpc.
    }
    \label{fig:lya_nb}
\end{figure*}

We constructed a pseudo-narrowband (NB) \Lya\ image by integrating the continuum-subtracted cube over 5719--5741~\AA\ (Extended Data Figure~\ref{fig:lya_nb}). Because of the astrometric mismatch between the MUSE cube and the JWST/NIRCam catalog frame, we did not adopt the JWST catalog position as the \Lya\ centroid. Instead, we defined the center from the brightest pixel in the \Lya\ NB image. We then measured the radial surface-brightness (SB) profile from the NB image in concentric circular annuli with a radial spacing of $0\farcs2$. Nearby contaminating sources were excluded using a segmentation map. 

To assess whether the observed \Lya\ emission is spatially extended, we compared its radial profile with the MUSE point-spread function (PSF). We model the PSF as a circular Moffat profile using the parameters from \cite{bacon_muse_2023}, with $\beta=2.8$ and
\[
{\rm FWHM}(\lambda)=a_1\lambda' + a_0,
\qquad
\lambda'=\frac{\lambda-4850}{9350-4850}-0.5,
\]
where $a_0=0.6179$ and $a_1=-0.1353$. At the effective wavelength of the \Lya\ narrowband image, $\lambda_{\rm eff}=5731$~\AA, this gives ${\rm FWHM}_{\rm PSF}=0.66\arcsec$. We compare the radial profiles of the \Lya\ emission and the MUSE PSF in Extended Data Figure~\ref{fig:lya_nb}. The observed \Lya\ profile is broader than the PSF, indicating spatially resolved emission extending to a projected radius of $\sim20$~kpc.

\begin{figure}
    \centering
    \includegraphics[width=0.9\linewidth]{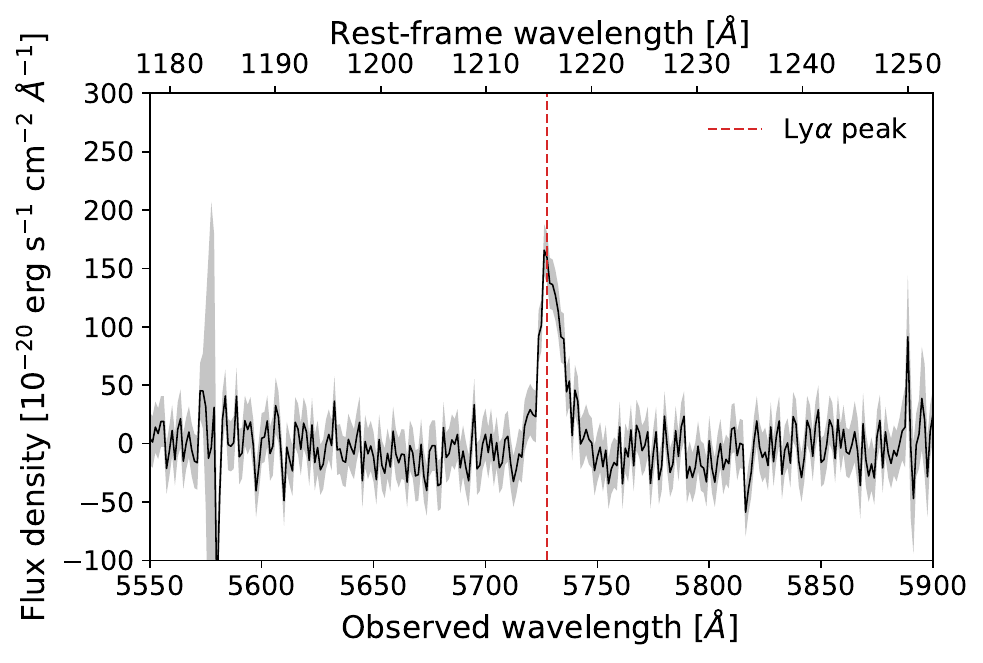}
    \caption{1D MUSE spectrum around \Lya. The spectrum was extracted from the continuum-subtracted emission-line cube after applying the spectral median-filter continuum subtraction. The extraction aperture is a circular aperture with radius $3\arcsec$, centered on the brightest pixel.}
    \label{fig:lya_1d}
\end{figure}

We extracted the one-dimensional \Lya\ spectrum from the continuum-subtracted MUSE emission-line cube by summing all spaxels within a circular aperture of radius $3\arcsec$, centered on the brightest pixel in the \Lya\ NB image. The masked regions lie outside this aperture and therefore do not affect the extracted spectrum. We show the extracted spectrum around \Lya\ in Extended Data Figure~\ref{fig:lya_1d}.

The integrated \Lya\ flux, luminosity, FWHM, and peak velocity offset were measured from the 1D spectrum. The uncertainty on $F_{\rm Ly\alpha}$ was obtained by summing the variance spectrum over the same wavelength range, and was propagated directly to $L_{\rm Ly\alpha}$.
The rest-frame \Lya\ equivalent width was computed using the \Lya\ flux from the continuum-subtracted cube and the continuum flux density measured from the original, non-continuum-subtracted cube using the same $3\arcsec$ aperture,
\begin{equation}
{\rm EW}_{\rm obs}({\rm Ly}\alpha)=
\frac{F_{\rm Ly\alpha}}{f_{\lambda,{\rm cont}}},
\end{equation}
and
\begin{equation}
{\rm EW}_{0}({\rm Ly}\alpha)=
\frac{{\rm EW}_{\rm obs}({\rm Ly}\alpha)}{1+z}.
\end{equation}
The continuum flux density was estimated from line-free spectral regions adjacent to \Lya. Its uncertainty was estimated from the aperture-summed variance spectrum in the same continuum windows and propagated together with the \Lya\ flux uncertainty. The UV absolute magnitude, $M_{\rm UV,1500}$, was measured from the original cube using a smaller $1\arcsec$-radius aperture and the continuum flux density around rest-frame 1500\,\AA{}, corresponding to $\lambda_{\rm obs}=1500(1+z)\,$\AA{}. The uncertainty in $M_{\rm UV,1500}$ was propagated from the variance spectrum in the same wavelength window.

The ratio of $L_{\rm{Ly}\alpha}$ to $L_{\mathrm{H}\alpha,\mathrm{narrow}}$ is $1.9\pm0.4$, which is below the Case~B recombination expectation of ${\rm{Ly}\alpha}/{\mathrm{H}\alpha} \sim 8.7$. This suppressed ratio likely reflects a low \Lya{} escape fraction caused by resonant scattering in neutral gas, which increases the effective path length and enhances the probability of dust absorption. The ratio may also depend on the gas geometry and covering fraction.

Because the extended \Lya\ halo has low surface brightness, we spatially rebinned the continuum-subtracted cube by $3\times3$ pixels before measuring the velocity moments. The velocity and velocity-dispersion maps were computed from the first and second flux-weighted moments of the \Lya\ profile. For each spatial bin, wavelength was converted to velocity relative to the systemic \Lya\ wavelength,
\begin{equation}
v(\lambda)=c\,\frac{\lambda-\lambda_{\rm Ly\alpha,obs}}{\lambda_{\rm Ly\alpha,obs}},
\end{equation}
where $\lambda_{\rm Ly\alpha,obs}=\lambda_{\rm Ly\alpha,rest}(1+z)$. The flux-weighted velocity and velocity dispersion were then calculated as
\begin{equation}
\langle v\rangle=
\frac{\sum_i v_i f_{\lambda,i}}{\sum_i f_{\lambda,i}},
\end{equation}
and
\begin{equation}
\sigma_v=
\left[
\frac{\sum_i (v_i-\langle v\rangle)^2 f_{\lambda,i}}
{\sum_i f_{\lambda,i}}
\right]^{1/2}.
\end{equation}
Here the sums are evaluated over spectral pixels within the adopted \Lya\ wavelength range. Negative flux values were clipped to zero when computing the moment maps to reduce noise-driven centroid errors, and spatial bins with ${\rm S/N}<2$ were masked in the final maps.

Uncertainties on the \Lya\ FWHM and peak velocity offset were estimated using Monte Carlo resampling. We generated perturbed realizations of the aperture-extracted continuum-subtracted spectrum using the aperture-summed STAT variance spectrum, remeasured the FWHM and peak wavelength for each realization, and adopted the 16th--84th percentile range as the statistical uncertainty. These uncertainties do not include systematic effects from the aperture choice, continuum-subtraction method, or adopted \Lya\ integration window.

\section*{Measuring host properties from image decomposition}

The source shows extended structure in the short-wavelength NIRCam bands. We therefore used JWST/NIRCam imaging to decompose the observed emission into a point source (i.e., the AGN) and an extended component (i.e., the host galaxy). The resulting host-galaxy photometry was then used to infer the stellar population properties, in particular the stellar mass.

We performed the image decomposition following the procedure described in \cite{ma_undermassive_2026}. We used NIRCam imaging in seven wide bands: F090W, F115W, F150W, F200W, F277W, F356W, and F444W. For each band, we extracted a $50\times50$ pixel cutout, corresponding to $1\farcs5\times1\farcs5$, centered on the source. Before the decomposition, we constructed masks for nearby contaminating sources using segmentation maps generated with \textsc{photutils}. These masks were used to exclude neighboring sources during the two-dimensional modeling.

We first used \textsc{GALFIT} \citep{peng_detailed_2002, peng_detailed_2010} to fit the F444W cutout with a single point source (PS) component. The resulting centroid was adopted as the AGN position and fixed across all bands. We then fit each band separately with \textsc{GALFIT}, using a PS component fixed at the AGN position and a S\'ersic component whose center was allowed to vary. The S\'ersic index was restricted to $0.7<n<8$, the effective radius to $0.1<R_{\rm e}<8$ pixels, and the axis ratio to $0.1<q<1$. For the adopted pixel scale, the effective-radius range corresponds to $0\farcs003$--$0\farcs24$, or approximately $0.02$--$1.68~{\rm kpc}$ at $z\simeq4$. These parameter ranges are consistent with those used in previous JWST image-decomposition studies \citep[e.g.,][]{zhuang_characterization_2024}.

Among the available bands, the F115W image provided the most stable and reliable S\'ersic fit, based on visual inspection of the residuals and comparison of the reduced $\chi^2$. We therefore adopted the best-fitting F115W S\'ersic structural parameters as the fiducial host-galaxy morphology: effective radius $R_{\rm e}=0\farcs094$, S\'ersic index $n=6.06$, axis ratio $q=0.78$, and position angle ${\rm PA}=15.35^\circ$. These structural parameters were fixed when constructing the host-galaxy spatial template in the other bands. Thus, the host morphology was assumed to be the same across all NIRCam bands, while the AGN and host fluxes were allowed to vary independently with wavelength.

To measure the component fluxes, we convolved the fixed S\'ersic host model with the corresponding PSF in each band and used it together with the PSF point-source model as two fixed spatial templates. We then fit each band independently with an MCMC procedure, allowing only the normalization of the AGN template, the normalization of the host template, and a constant background term to vary. This yields the flux contribution of the unresolved AGN and the extended host component in each band. We then subtracted the best-fitting AGN component from the original images to construct AGN-subtracted images.

We measured the extended-emission fluxes from the AGN-subtracted images using aperture photometry. Total fluxes were measured from the original images using the same aperture definition. We adopted a circular aperture with radius $0\farcs45$ and estimated the local background in a sky annulus with inner and outer radii of $0\farcs60$ and $0\farcs75$, respectively. Aperture corrections were calculated separately for each band using the corresponding PSF.

\begin{figure*}
    \centering
    \includegraphics[width=0.8\linewidth]{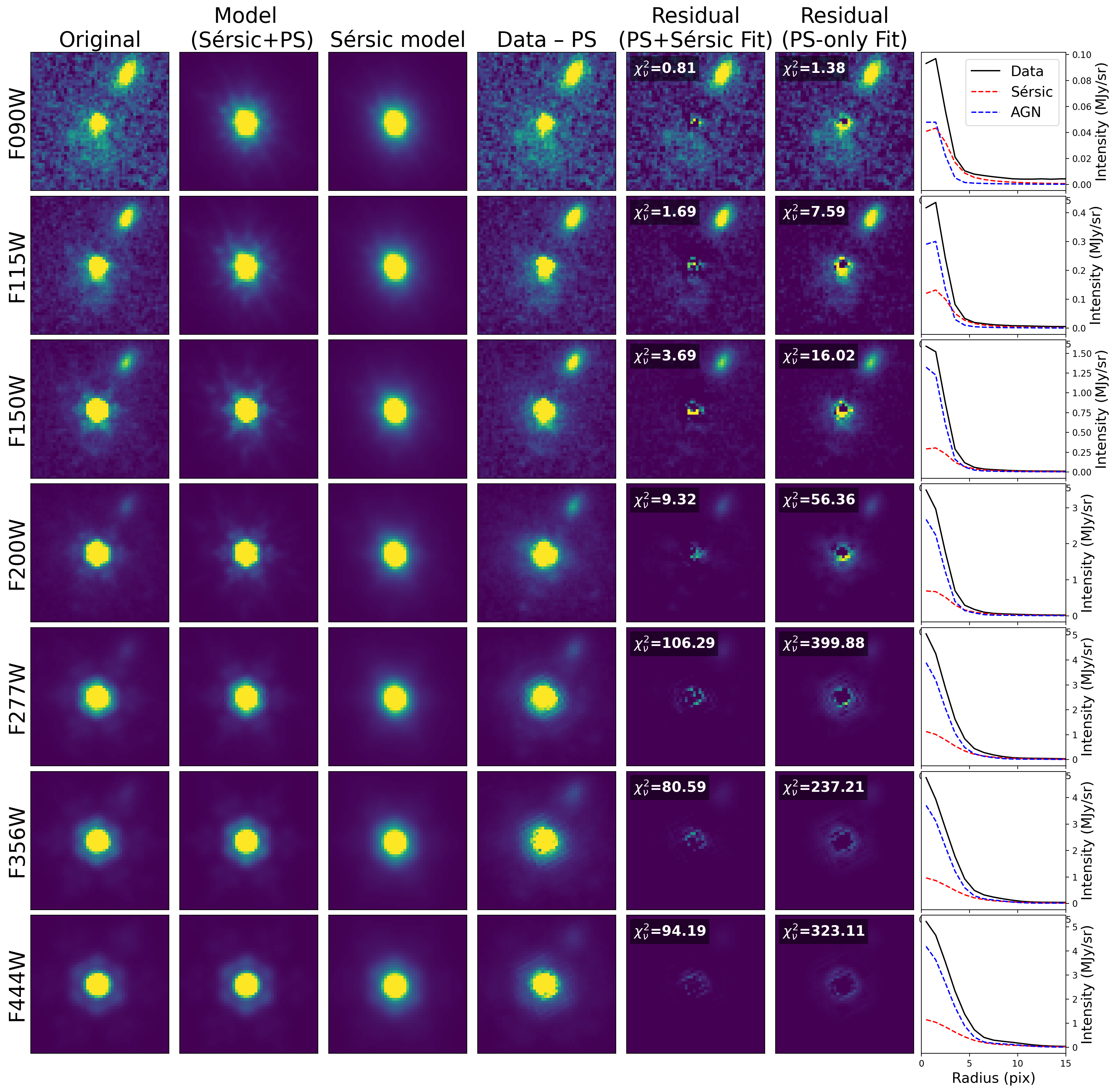}
    \includegraphics[width=0.8\linewidth]{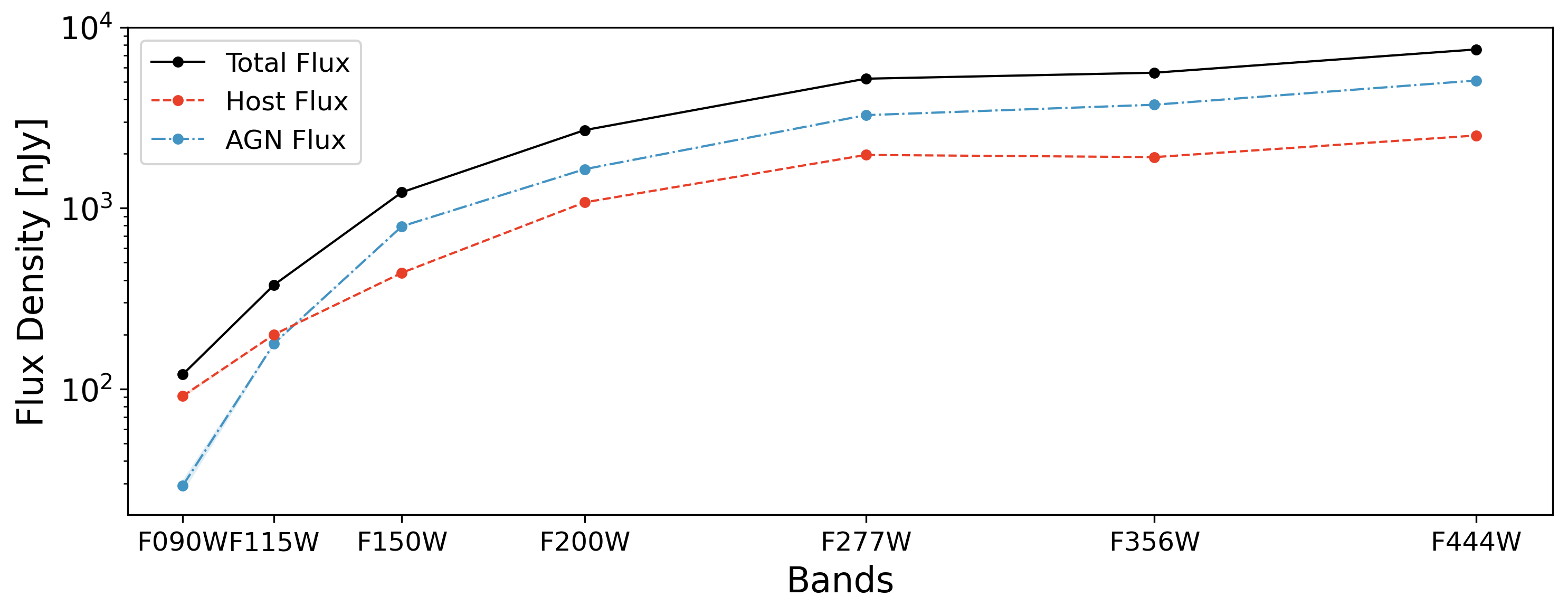}
    \caption{AGN-Host decomposition of the Hatchling. The cutout image size is $1.5^{\prime\prime}\times 1.5^{\prime\prime}$. The first column displays the original image in each band. The second column shows the two-component model (PSF + S\'ersic). The third column shows the S\'ersic (i.e., host-galaxy) model. The fourth column shows the AGN-subtracted images, which show residual extended emission in the short-wavelength bands as a signal of host detection.  The fifth column shows the residual images after subtracting the PSF$+$S\'ersic model, with the fitting reduced $\chi^2$ denoted in each image. The sixth column shows the residual images when we attempt to fit the source with one single PSF model, with the fitting reduced $\chi^2$ denoted. The last column displays the radial brightness profile of the original data (the black solid line), S\'ersic model (the red dashed line), and PSF AGN model (the blue dashed line). The pixel scale is 0.03$\arcsec{}$ per pixel. 
    \textbf{Bottom panel:} SED of the Hatchling. The black solid line shows the total flux, and the blue dash-dotted line shows the flux of the point-source AGN model.  The red dashed line shows the host flux measured from the AGN-subtracted images using aperture photometry.}
    \label{fig:image_decomposition}
\end{figure*}

\begin{figure*}[!ht]
    \centering
    \includegraphics[width=0.8\linewidth]{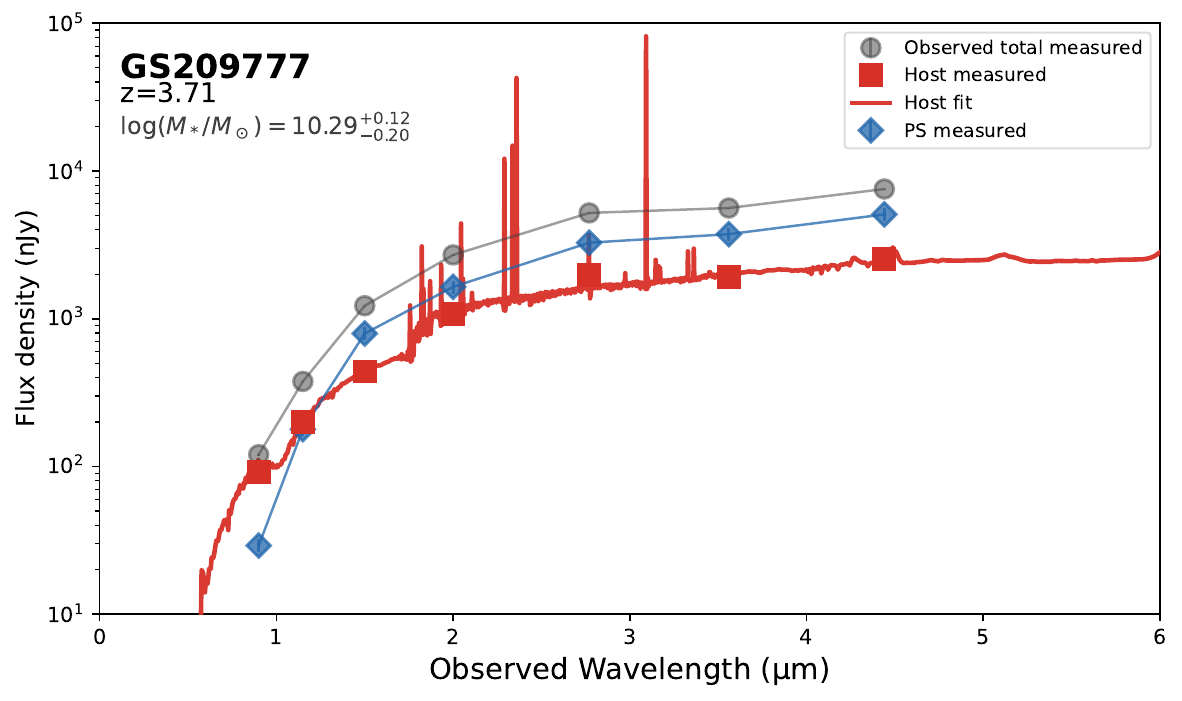}
    \caption{SED fits for the Hatchling (GS209777). Gray circles show the observed total photometry, and blue diamonds show the decomposed
point-source photometry. Solid red lines show host-galaxy SED fits based on image decomposition.}
    \label{fig:sed_fitting}
\end{figure*}

The resulting image decomposition is shown in Extended Data Figure~\ref{fig:image_decomposition}. We fitted the AGN-subtracted host-galaxy photometry with \textsc{Prospector} \citep{johnson_stellar_2021}, following the setup in \cite{zhu_smiles_2025}. The model assumes a Chabrier initial mass function and includes stellar-population, dust-attenuation, metallicity, and nebular-emission components. We used a non-parametric star-formation history with seven age bins and a continuity prior \citep{leja_how_2019}; the bin edges are logarithmically spaced from $10^{7.13}$ yr to $0.9\,t_{\rm univ}$, with a final bin extending to $t_{\rm univ}$, where $t_{\rm univ}$ is the age of the Universe at the source redshift. We used only the decomposed NIRCam host photometry in the fit; the MIRI bands were not included because the strong mid-infrared emission and MIRI-based AGN identification indicate a substantial AGN contribution \citep{lyu_active_2024}. The SED-fitting result is shown in Extended Data Figure~\ref{fig:sed_fitting}. From the posterior distribution, we infer $\log(M_\star/M_\odot)=10.3^{+0.1}_{-0.2}$ and a recent star-formation rate averaged over the past 30 Myr of $\mathrm{SFR}_{30}=54^{+45}_{-33}\,$\Msunyr{}.

Our inferred stellar mass is lower than the value reported by \cite{hoshi_evolutionary_2025}, who found $\log(M_\star/M_\odot)=11.02\pm0.02$ for the same source, by $\simeq0.7$ dex, or a factor of $\sim5$. They used CIGALE to estimate masses from the SEDs and we used PROSPECTOR; the masses returned for these two approaches for z $\sim$ 3 have been shown to be in good agreement \cite{pacifici_art_2023}. This difference is thus likely driven in part by the treatment of the PSF in the image decomposition. We use the JADES model PSFs, which are generated for each subregion, filter, and mosaic position angle following methods developed in \cite{ji_jades_2024} and described in \cite{johnson_jwst_2026,robertson_jwst_2026}. These model PSFs include the broadening of the theoretical PSF introduced by drizzling and therefore provide an empirical match to the final mosaics used for the decomposition. In contrast, \cite{hoshi_evolutionary_2025} used theoretical WebbPSF models, which are generally sharper than the effective PSF in the drizzled mosaics. For a compact, quasar-dominated source such as the Hatchling, adopting a sharper PSF can leave excess central light in the extended component and therefore increase the inferred host flux and stellar mass. 
 
A qualitative difference in the two results is that we find the galaxy to be fainter than the quasar in all bands redder than F115W, whereas \citep{hoshi_evolutionary_2025} find the galaxy to be brighter than the quasar. This latter result is difficult to reconcile with the rest-frame near-infrared  MIRI photometry being well fit by a standard Elvis quasar template \cite{elvis_atlas_1994}. That is, the strong mid-infrared output also supports a substantial quasar contribution to the observed emission, consistent with our lower host-galaxy stellar mass after subtracting the unresolved component.

\section*{Surrounding environment}
\begin{figure*}
    \centering
    \includegraphics[width=0.8\linewidth]{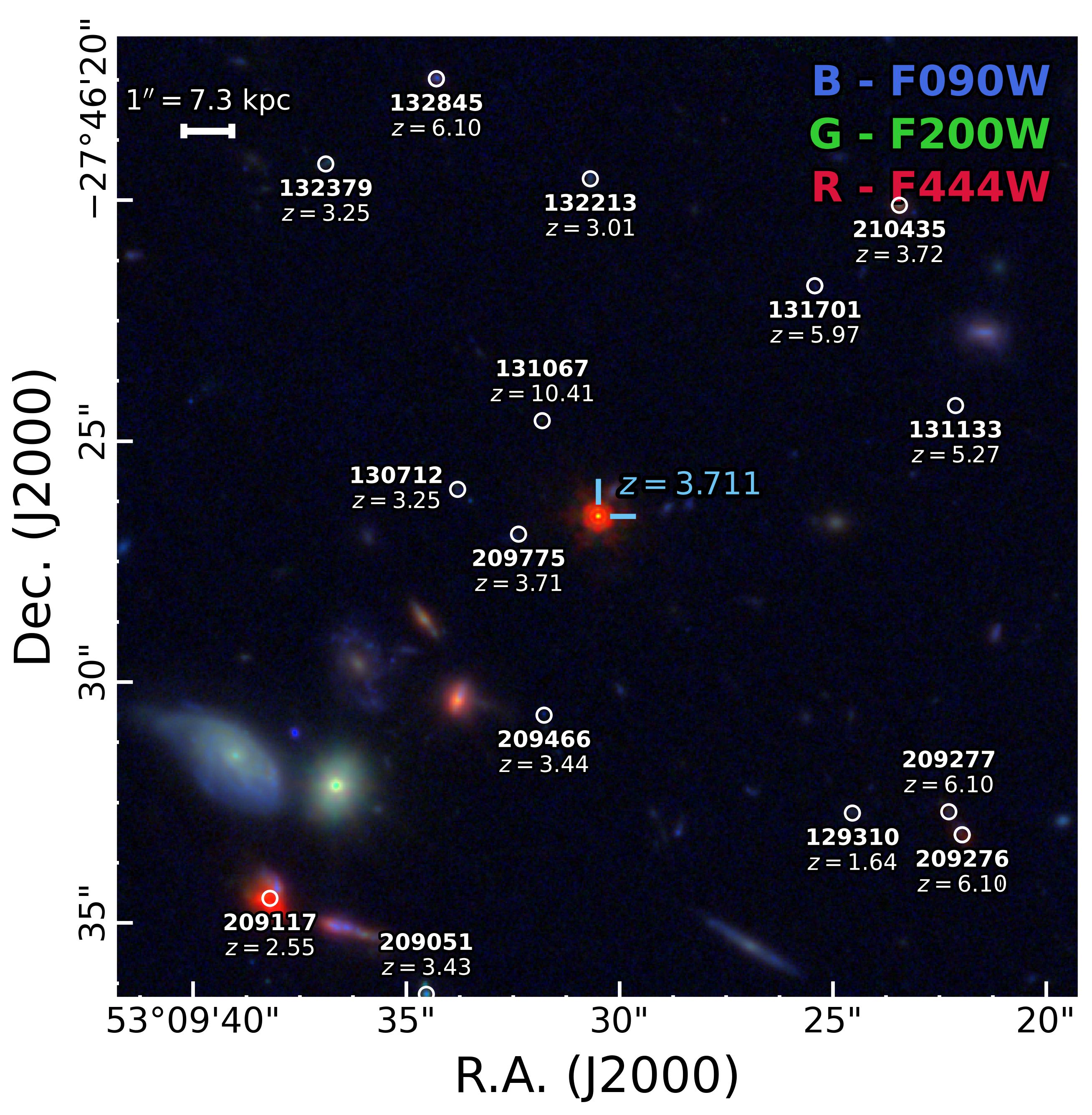}
    \caption{False-color JWST/NIRCam image of the Hatchling and nearby galaxies. The cutout size is $20\farcs0 \times 20\farcs0$. Nearby sources are labeled with their JADES NIRCam IDs and spectroscopic redshifts from the NIRSpec catalog. GS209775 and GS210435 have redshifts close to that of the Hatchling, indicating that these sources may trace a local overdensity around the Hatchling. }
    \label{fig:environment}
\end{figure*}

To examine the local environment of the Hatchling, we inspect the spectroscopic redshifts of sources within a ($20\farcs0 \times 20\farcs0$) region centered on the target. As shown in Extended Data Figure~\ref{fig:environment}, we identify two nearby sources, GS209775 at $z = 3.71$ and GS210435 at $z = 3.72$, that are close to the Hatchling in both projected position and redshift. They lie at projected separations of approximately 12 and 66 proper kpc, respectively. The presence of these nearby galaxies may indicate that the Hatchling resides in a locally overdense environment.

This local structure may also be connected to a larger-scale overdensity in the GOODS-S field. Sun et al. \cite{sun_jades_2026} identify a prominent $z\simeq3.69$ protocluster in the JADES/JOF region, including ALESS9.1 and the Cosmic Rose, and suggest that its filamentary structure extends northeast toward the HUDF field. The Hatchling may therefore be associated with this broader large-scale structure, although a direct physical connection remains tentative without a more complete spectroscopic census across the field. If physically associated, such an environment could promote galaxy interactions, mergers, or gas accretion and may contribute to the continued buildup of stellar mass in the Hatchling \cite{stone_z_2025}.

\bigskip

\section*{Data Availability}
This work is based on observations made with the NASA/ESA/CSA James Webb Space Telescope. The data were obtained from the Mikulski Archive for Space Telescopes at the Space Telescope Science Institute, which is operated by the Association of Universities for Research in Astronomy, Inc., under NASA contract NAS 5-03127 for JWST. All the JWST data used in this paper can be found in MAST: \url{https://doi.org/10.17909/8tdj-8n28}.

\section*{Code Availability}
Analysis was performed using {\tt Astropy} \citep{astropy_collaboration_astropy_2022}, {\tt GALFIT} \citep{peng_detailed_2002, peng_detailed_2010}, and {\tt PyQSOFit} \citep{guo_pyqsofit_2018, shen_sloan_2019}.


\section*{References}
\bibliography{references_3}{}

\section*{Correspondence}
Correspondence and requests for materials should be addressed to Zheng Ma and Yongda Zhu.

\section*{Acknowledgments}
We thank Masami Ouchi and his group members for helpful discussions.
ZM, YZ, ZJ, GHR, MJR, EE and CNAW acknowledge support from the NIRCam Science Team contract to the University of Arizona, NAS5-02105. YZ is also supported by JWST Program \#6434. Support for program \#6434 was provided by NASA through a grant from the Space Telescope Science Institute, which is operated by the Association of Universities for Research in Astronomy, Inc., under NASA contract NAS 5-03127. AJB acknowledges funding from the “FirstGalaxies” Advanced Grant from the European Research Council (ERC) under the European Union’s Horizon 2020 research and innovation program (Grant agreement No. 789056). FDE, IJ and RM acknowledge support by the Science and Technology Facilities Council (STFC), by the ERC through Advanced Grant 695671 ``QUENCH'', and by the UKRI Frontier Research grant RISEandFALL. IJ acknowledges support by the Huo Family Foundation through a P.C. Ho PhD Studentship. RM also acknowledges funding from a research professorship from the Royal Society.

We respectfully acknowledge that the University of Arizona is on the land and territories of Indigenous peoples. Today, Arizona is home to 22 federally recognized tribes, with Tucson being home to the O’odham and the Yaqui. The university strives to build sustainable relationships with sovereign Native Nations and Indigenous communities through education offerings, partnerships, and community service.

\section*{Author Contributions}
ZM led the analysis, including the NIRSpec and MUSE spectral measurements, image decomposition, and physical interpretation. YZ led the discussion and interpretation of the source. ZM and YZ wrote the first draft of the manuscript. ZJ contributed to the target selection, identified the source as an object of interest in archival NIRSpec data, and initiated discussion of its unusual spectral features. EE, MJR and XF contributed to the overall scientific framing and presentation. JL contributed to the AGN identification using multiwavelength data and GHR helped with the science interpretation. YS contributed to the identification and interpretation of the Na\,{\sc d} outflow. All authors discussed the results and commented on the manuscript. AJB was involved in the target selection, planning and data release of the JADES NIRSpec spectra.

\section*{Competing interests}
The authors declare no competing interests.

\end{document}